\documentclass[twocolumn, trackchanges]{aastex701}

\usepackage{gensymb}
\usepackage{amsmath}
\usepackage{enumitem}
\usepackage{flafter}
\usepackage{xspace}
\usepackage{rotating}
\usepackage{tablefootnote}

\usepackage{graphicx}
\usepackage{xcolor}
\usepackage{booktabs}

\definecolor{darkgreen}{RGB}{0,100,0}

\newcommand{\avd}{AT~2019avd\xspace}

\newcommand{\msunyr}{M_{\odot}~ {\rm yr^{-1}}}

\shorttitle{A Bow‑Shock Interpretation for \avd}
\shortauthors{Wei et al.}

\begin{document}

\title{\centering Radio Emission with Dust: A Bow-Shock Interpretation for the TDE Candidate \avd}

\author[0009-0005-5669-8465]{Shuyuan Wei}
\affiliation{School of Astronomy and Space Sciences, University of Chinese Academy of Sciences, Beijing 100049, China}
\email{weishuyuan24@mails.ucas.ac.cn}

\author[0000-0003-3207-5237]{Yanan Wang}
\altaffiliation{E-mail: wangyn@bao.ac.cn}
\email{wangyn@bao.ac.cn}
\affiliation{National Astronomical Observatories, Chinese Academy of Sciences, 20A Datun Road, Beijing 100101, China}

\author[0000-0002-0092-7944]{Guobin Mou}
\altaffiliation{E-mail: gbmou@njnu.edu.cn}
\email{gbmou@njnu.edu.cn}
\affiliation{Department of Physics and Institute of Theoretical Physics, Nanjing Normal University, Nanjing 210023, China}

\author[0000-0003-3454-6522]{Linhui Wu}
\email{wulinhui@shao.ac.cn}
\affiliation {Shanghai Astronomical Observatory, Chinese Academy of Sciences, 80 Nandan Road, Shanghai 200030, China}
\author[0000-0002-5761-2417]{Santiago del Palacio}
\email{santiago.delpalacio@chalmers.se}
\affiliation{Department of Space, Earth and Environment, Chalmers University of Technology, SE-412 96 Gothenburg, Sweden}

\author[0000-0002-1824-0411]{Ranieri D. Baldi}
\email{ranieri.baldi@inaf.it}
\affiliation{INAF - Istituto di Radioastronomia, Via P. Gobetti 101, I-40129 Bologna, Italy}

\author[0000-0002-4439-5580]{Xiaolong Yang}
\email{yangxl@shao.ac.cn}
\affiliation {Shanghai Astronomical Observatory, Chinese Academy of Sciences, 80 Nandan Road, Shanghai 200030, China}

\author[0000-0003-3250-2876]{Yang Huang}
\email{huangyang@ucas.ac.cn}
\affiliation{School of Astronomy and Space Sciences, University of Chinese Academy of Sciences, Beijing 100049, China}
\affiliation{National Astronomical Observatories, Chinese Academy of Sciences, 20A Datun Road, Beijing 100101, China}

\begin{abstract}
Forward shocks produced by interactions between a single ejected blob and the circumnuclear medium (CNM) around a supermassive black hole are widely invoked to explain radio emission from tidal disruption events (TDEs). However, recent observational evidence such as rapidly rising radio emission, double-peaked broadband spectral energy distributions, and declining shock energies at late times poses challenges to this picture. Alternative scenarios have therefore been proposed, including bow shocks arising from outflows colliding with dense clouds, the production of new ejecta at late times, and variations in microphysical parameters.
To investigate the roles of these factors, we analyze the long-lived radio flare of the TDE candidate \avd, for which dust has been identified through mid-infrared observations, using a radio dataset spanning more than five years. We find that a bow-shock scenario can account for the majority of the observed radio emission, while a forward shock, or an additional bow shock involving different clouds, may begin to dominate at late times. However, a forward-shock scenario remains viable when allowing for evolving microphysical parameters and more complex CNM structures. We discuss the remaining tensions in both the outflow-CNM and outflow-cloud interaction scenarios.

\end{abstract}

\keywords{\uat{High energy astrophysics}{739} --- \uat{Supermassive black holes}{1663} --- \uat{Radio transient sources}{2008} --- \uat{Radio continuum emission}{1340}}


\section{Introduction} \label{sec:intro}
Tidal disruption events (TDEs) are nuclear transients that occur when a star is tidally shredded upon passing too close to a supermassive black hole (SMBH) \citep{Hills1975, Rees1988}. Their emission can temporarily illuminate dormant SMBHs and the unresolved nuclear regions of distant galaxies on sub-parsec scales. Characterizing the circumnuclear medium (CNM), including gas and dust, requires indirect diagnostic methods. For example, extreme coronal-line emitters probe the properties of circumnuclear gas around SMBHs (e.g. \citealt{Newsome2024}), while mid-infrared (IR) emission traces the reverberation radii of dust heated by optical to soft X-ray flares from TDEs (e.g. \citealt{Jiang2016,Lu2016,Velzen2016}). 
Radio emission in TDEs is generally attributed to synchrotron radiation from external shocks produced as a TDE-driven outflow or jet interacts with the CNM (see \citealt{Alexander2020} for a review). ASASSN-14li \citep{Pasham2018} and AT~2020afhd \citep{Wang2025} are two reported exceptions, with their radio emission instead linked directly to jet activity through strong X-ray--radio correlations. In contrast to observations at other wavelengths, which primarily probe radiative energy output, radio observations uniquely constrain the properties of the ejected material.

Radio TDEs exhibit a wide range of radio luminosities, spanning from $10^{37}$ to $10^{42}\rm \,erg\,s^{-1}$, with substantial diversity in their durations, onset times, evolution of the spectral energy distribution \citep[SED;][]{Alexander2020}, and short-term variability \citep{Wang2025}. 
To date, four events with luminous radio emission, $\nu L_\nu\gtrsim10^{40}\rm erg\,s^{-1}$, have been attributed to on-axis relativistic jets, i.e., Swift~J1644+57 \citep{Bloom2011,Zauderer2011,Berger2012,Eftekhari2018}, Swift~J2058+05 \citep{Cenko2012}, Swift~J1112.2-8238 \citep{Brown2015}, and AT2022cmc \citep{Andreoni2022,Pasham2023,Rhodes2023}. By contrast, many optically discovered TDEs exhibit substantially fainter radio emission, with $\nu L_\nu\lesssim10^{39}\rm erg\,s^{-1}$. The physical mechanisms powering this population remain uncertain. Proposed scenarios include off-axis jets (e.g., AT~2018hyz; \citealt{Cendes2022,Matsumoto2023,Sfaradi2024}), internal shocks within jets (e.g., ASASSN-14li; \citealt{Pasham2018}), non-relativistic winds (e.g., AT~2019dsg; \citealt{Cendes2021a}; AT~2020opy; \citealt{Goodwin2023a}; AT~2020zso; \citealt{Christy2025}; AT~2024tvd; \citealt{Sfaradi2025}), collision-induced outflows (e.g., AT~2019azh; \citealt{Goodwin2022}; AT~2020vwl; \citealt{Goodwin2023b}), and unbound tidal debris streams (theoretically; \citealt{Krolik2016,Yalinewich2019}; and as an alternative for AT~2020vwl; \citealt{Goodwin2023b}).

While most models invoke a forward shock produced as an outflow propagates into the hot, diffuse circumnuclear medium (CNM; \citealt{Alexander2020}), bow shocks generated by interactions between the outflow and dense clouds have also been proposed as an alternative \citep{mou2021,mou2022}.
The main distinction between these scenarios lies in the physical interpretation of the radio-inferred parameters. 
In the forward-shock picture, the inferred radius corresponds to the location of the expanding shock front, and the inferred density represents the CNM density at that radius, enabling constraints on the radial density profile under assumptions about the outflow geometry and microphysics. In the bow-shock picture, by contrast, the radio emission originates from shocks wrapping around dense clouds. The inferred quantities then characterize the bow shock properties themselves: the inferred radius corresponds to their characteristic scale, which is expected to be comparable to the cloud radius, whereas the inferred density traces that of the incident outflow rather than the ambient CNM. This latter scenario has been proposed to explain the delayed and rapidly rising radio emission of AT~2018cqh, and hydrodynamic calculations have shown that it can reproduce the observed radio evolution \citep{Yang2025}. Additional observations also suggest that a single-component forward-shock model may not always provide a complete description. For example, the transient double-peaked radio SED of AT~2020zso provides evidence for multiple physically distinct outflow components, while the interpretation of the late-time radio evolution of AT~2021sdu is complicated by possible host-galaxy contamination \citep{Christy2025}. More complex outflow structures, multiple ejecta episodes, environmental inhomogeneities, or evolving microphysical parameters may therefore be required in some events. Identifying the physical origin and geometry of the radio-emitting shocks is essential for a robust interpretation of the inferred outflow and CNM properties.


\avd\ is a nuclear transient characterized by two consecutive flaring episodes in the optical and IR bands \citep{Malyali2021,Wang2023}. A growing number of studies have shown that TDEs can exhibit optical re-brightening \citep{Yao2023}. In \avd, the X-ray decline lagged the optical and UV fading by approximately 80 days, in contrast to the behavior commonly observed in AGN. \citet{Malyali2021} also reported Bowen fluorescence features and high-ionization coronal lines in the optical spectra of \avd. Radio emission coincident with the second optical flare and the X-ray brightening was subsequently detected \citep{Wang2023}. Very Large Array (VLA) and Very Long Baseline Array (VLBA) observations revealed the emergence of a compact radio source with a non-thermal spectrum, interpreted as emission from an evolving ejecta component associated with a change in the black-hole accretion state \citep{Wang2023}. 
\cite{Goodwin2025} further showed that the inferred outflow properties of \avd\ are broadly consistent with those of other radio TDEs. Taken together, these properties favor a TDE interpretation, although the possibility that \avd\ represents a new class of nuclear transient cannot yet be excluded.

The mid-IR emission from \avd is consistent with a dust echo from material located at a distance of approximately 0.03\,pc from the SMBH. The long-lived radio emission, together with the presence of circumnuclear dust, makes \avd an ideal laboratory for probing the temporal evolution of outflows, distinguishing their physical origins, and assessing the relative roles of forward shocks and bow shocks.

In this work, we present a comprehensive analysis of the radio properties of \avd, utilizing data collected over five years. The paper is structured as follows. Sect.~\ref{sec:observation} describes the radio observations and data reduction; Sect.~\ref{sec:Results} presents the multi-epoch spectral fits, and tests (i) an outflow–CNM forward-shock interpretation using analytical prescriptions and (ii) an outflow-cloud bow-shock interpretation using hydrodynamical simulations; Sect.~\ref{sec:disscussion} discusses the outflow origin under outflow-CNM scenario and potential tensions with both the outflow-CNM and outflow-cloud scenarios; and Sect.~\ref{sec:conclusion} summarizes our conclusions. 


\section{Observations and data reduction}\label{sec:observation}
We present the observation log for \avd\ in Table~\ref{tab:radio_obs}. The log incorporates data from \cite{Wang2023}, which reports some VLA and VLBA flux measurements, as well as our own observational dataset.

In this work, uncertainties and upper/lower limits are quoted at the 1$\sigma$ and 3$\sigma$ confidence levels, respectively. We adopt a redshift of 0.028 from the Transient Name Server\footnote{\url{https://www.wis-tns.org/object/2019avd}}, corresponding to a luminosity distance of $D=130$\,Mpc \citep{Wang2023}. The black hole mass is taken to be $M_{\rm BH}=10^{6.3}\,M_{\odot}$ from \cite{Malyali2021}. 

\subsection{VLA} 
We obtained three radio observations of \avd\ with the VLA (program IDs: 23B-301 and 24A-434; PI: Y. Wang). These observations cover the L, S, C, X, and K bands (see more details in Table.~\ref{tab:radio_obs}).

The data were reduced in the Common Astronomy Software Applications package (CASA v6.6.6; \citealt{CASA}). We applied the VLA CASA calibration pipeline (v2025.1.0.32) for standard calibration and automated flagging, followed by automated self-calibration using the VLA imaging pipeline task \textsc{hif\_selfcal}. Calibrated visibilities were imaged with \textsc{tclean} into continuum images in frequency bins, using \textsc{auto-multithresh} to guide deconvolution. The resulting images typically reach a residual rms of $\sim100\,\mu$Jy\,$\mathrm{beam}^{-1}$. 

Unless explicitly mentioned, flux densities were measured by fitting a single Gaussian at the position of \avd\ in the cleaned images using \textsc{imfit}. For measurements obtained with all facilities, the uncertainties were computed by adding the statistical error in quadrature with a 5\% systematic term to account for the absolute flux scale.


\subsection{ATCA}
We conducted four epochs of observations of \avd\ using the Australia Telescope Compact Array (ATCA) as part of programs C3615, C3662, and C3755 (PI: Y. Wang). Each epoch included observations in the C/X-band centered at 5.5\,GHz and 9\,GHz (with a 2\,GHz bandwidth per window) and in the L-band centered at 2.1\,GHz (with a 2\,GHz bandwidth). For all ATCA observations, 0823$-$500 was used as the flux and bandpass calibrator, and 0823+033 as the phase calibrator.

The data were reduced in CASA following standard procedures, and images were produced using the task \textsc{tclean}. In the L band, significant sidelobe contamination from nearby bright sources was identified. To improve the dynamic range, we performed two rounds of phase-only self-calibration on the L-band target visibilities. 

In the C-band observation of epoch~9, the relatively large synthesized beam resulted in mild blending with a nearby source. We therefore employed two CLEAN masks, i.e., one centered on \avd\ and one on the nearby source, during deconvolution, and adopted the integrated flux density measured from the \textsc{tclean} CLEAN-component \texttt{model} image within an aperture matched to the restoring beam and centered on the position of \avd. The corresponding statistical uncertainty was taken to be the local rms noise measured in a nearby source-free region of the corresponding restored image.

\subsection{VLASS}
We analyzed the Karl G.~Jansky Very Large Array Sky Survey (VLASS; \citealt{VLASS}) epoch~3 data covering the position of \avd\ using the user-defined VLASS imaging scripts \citep{VLASS_Scripts}. We first applied the VLASS Single-Epoch Imaging Pipeline (SEIP) to perform phase-only self-calibration on the standard pipeline-calibrated measurement sets. To constrain the spectral properties of the source, we subsequently ran the Coarse-Cube Imaging Pipeline (CCIP), producing images in two spectral sub-bands centered at 2.5 and 3.5\,GHz, each with a bandwidth of 1\,GHz. 

The flux densities for each epoch are listed in Table~\ref{tab:radio_obs}. The target remained unresolved throughout our observations.

\section{Results}\label{sec:Results}
\begin{figure*}[htb!]
\centering
    \includegraphics[width=0.95\textwidth]{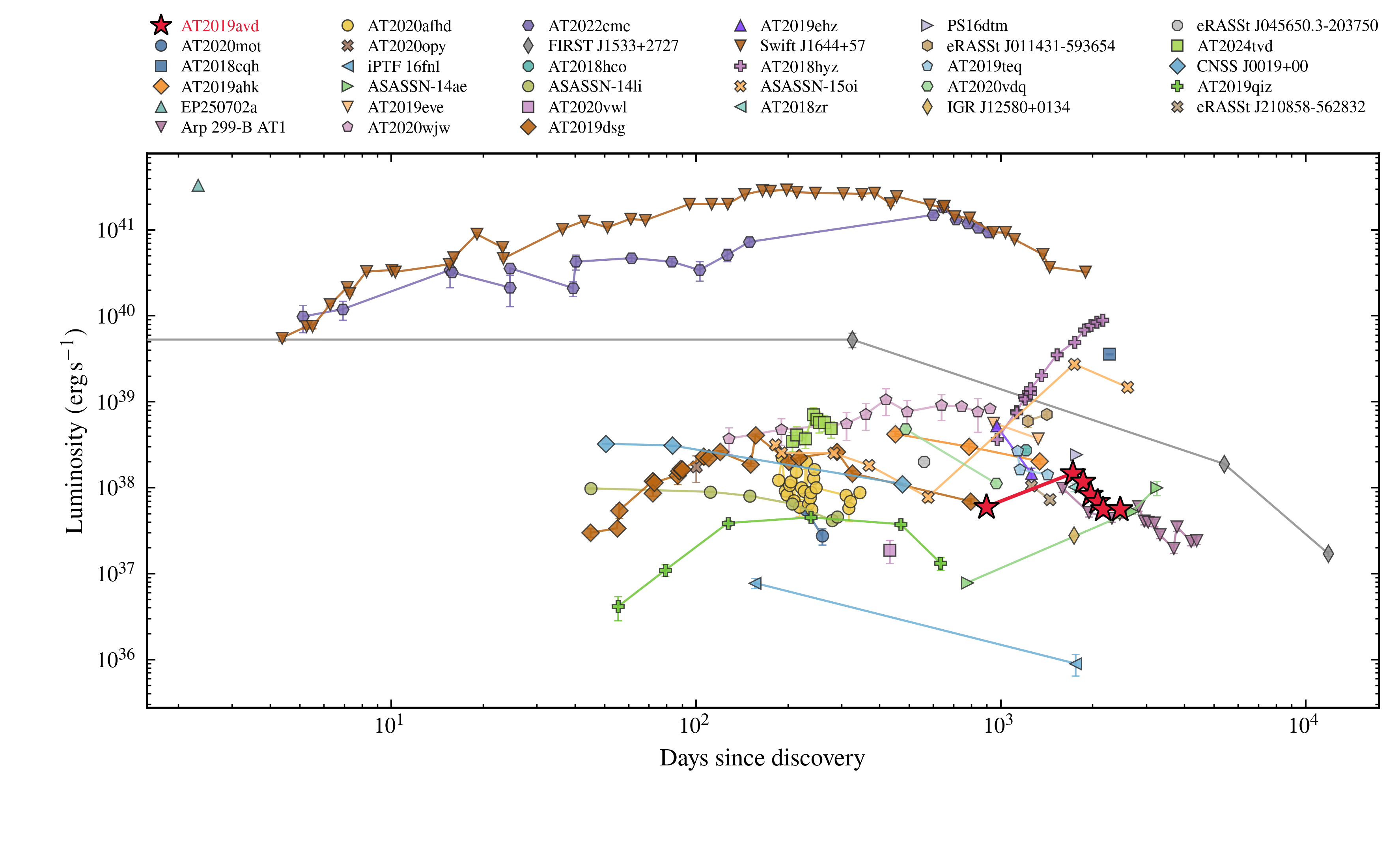} 
    \caption{The 5\,GHz light curves of radio-detected TDEs selected from the sample of \citealt{Zhou2026}. We retain only measurements obtained within $4.8\leq\nu\leq5.2\,\mathrm{GHz}$, and define the time offsets relative to the adopted discovery date of each source. The measurements of \avd are presented in this work. The comparison sources and the references for their radio observations are summarized in Table~\ref{tab:radio_tde_sample}.}

    \label{fig:lightcurve_TDEs}
\end{figure*}

To place the radio evolution of \avd in the context of other radio-detected TDEs, Fig.~\ref{fig:lightcurve_TDEs} compares its 5\,GHz luminosity with those of a selected sample. \avd exhibits a delayed rise, reaches its peak radio luminosity at 1445\,days after discovery, and subsequently fades rapidly. Its radio luminosity remains substantially below those of the canonical on-axis jetted TDEs.

\subsection{Spectral fitting}\label{subsec:spec_fit}

\begin{figure*}[h!]
\centering
\includegraphics[width=0.95\textwidth]{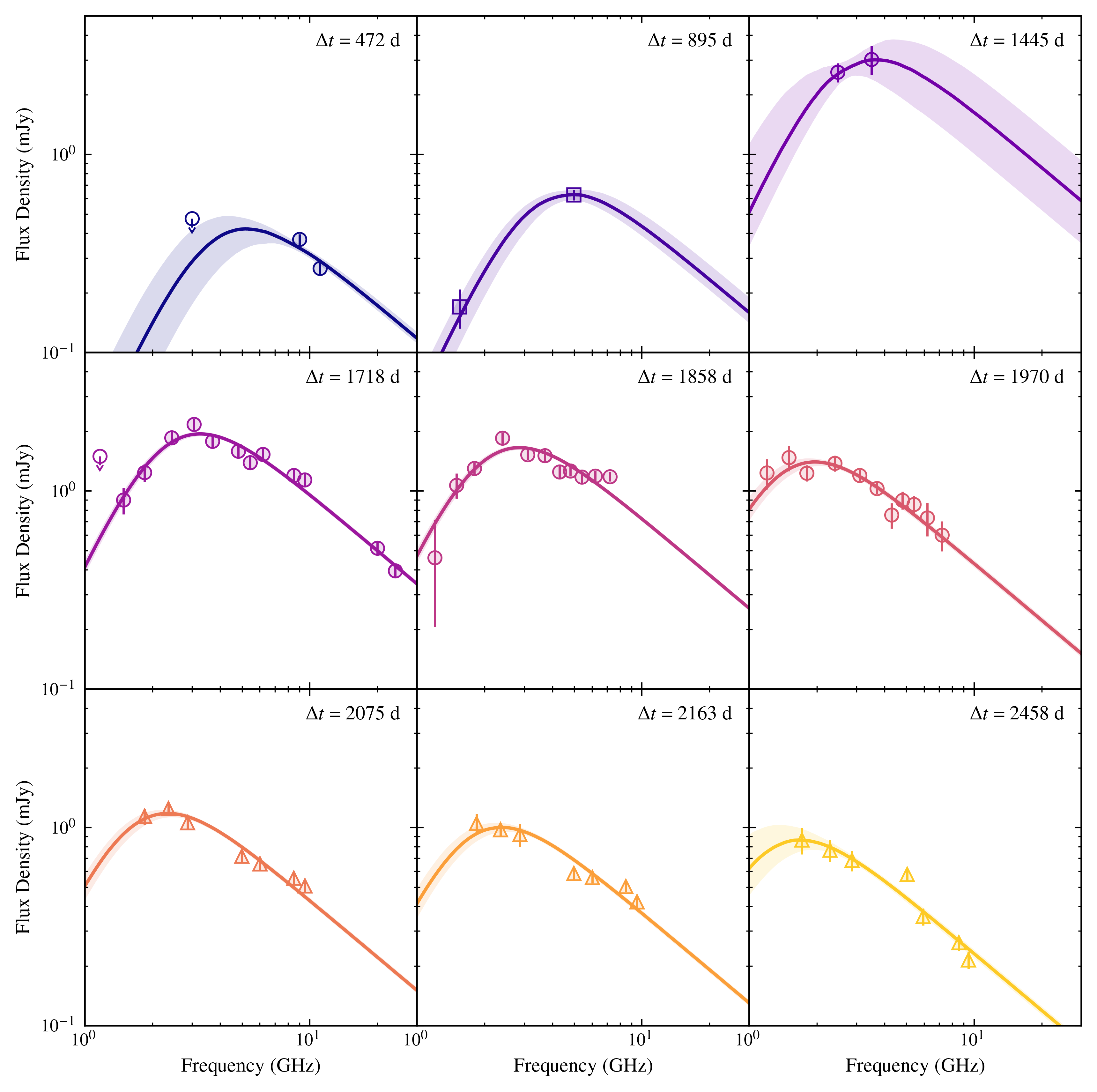}
\caption{Multi-epoch SEDs. Each panel corresponds to a single epoch and is labeled by the time since optical discovery, $\Delta t \equiv t_\mathrm{obs}-t_\mathrm{d}$, where $t_\mathrm{d}=58523.23$ denotes the time of the optical discovery. Circular, square, and triangular markers denote observations from the VLA, VLBA, and ATCA, respectively, while open symbols with downward arrows represent $3\sigma$ upper limits. The solid lines are the best-fit solutions using the phenomenological model from Eq.~\ref{eq:model_ssa}, and the shaded regions represent the $1\sigma$ confidence intervals of the model fit at each frequency.
\label{fig:sed_fitting}}
\end{figure*}

We fitted the radio SED at each epoch using the synchrotron self-absorbed (SSA) spectrum described by \citet{granot02}, adopting the spectral regime $\nu_\mathrm{m}<\nu_\mathrm{a}<\nu_\mathrm{c}$ (where $\nu_\mathrm{m}$ is the synchrotron frequency corresponding to the minimum electron Lorentz factor, $\nu_\mathrm{a}$ is the SSA frequency, and $\nu_\mathrm{c}$ is the synchrotron cooling frequency):
\begin{equation}\label{eq:model_ssa}
  F(\nu)=F_\mathrm{norm} \times \left(\frac{\nu}{5\,\mathrm{GHz}}\right)^{b_2} \times \left[1+\left(\frac{\nu}{\nu_a}\right)^{s_2(b_2-b_3)}\right]^{-\frac{1}{s_2}}.   
\end{equation}
Here, $F_\mathrm{norm}$ is the normalization flux density, and $p$ is the electron energy index for a power‑law distribution $N(\gamma_{\mathrm{e}})\propto \gamma_{\mathrm{e}}^{-p}$ ($\gamma_\mathrm{e}\geq\gamma_{\mathrm{m}}$, where $\gamma_{\mathrm{m}}$ is the minimum Lorentz factor). 
The spectral slopes are fixed as $b_2=2.5$ (SSA slope) and $b_3=\frac{1-p}{2}$ (optically thin slope), with the break smoothness parameter given by $s_2=1.47-0.21p$. 

We performed Bayesian inference using \texttt{bilby}
\citep{bilby_paper,bilby_doi} with the \texttt{dynesty} sampler
\citep{2020MNRAS.493.3132S,koposov_dynesty_zenodo}, employing 1000 live points and terminating the sampling when the estimated remaining log-evidence satisfied $\Delta\ln \mathcal{Z} < 0.1$. We adopted uniform priors of $\nu_\mathrm{a} \in [0.1, 9.0]\,\mathrm{GHz}$ and $F_\mathrm{norm} \in [10^{1}, 10^{6}]\,\mu\mathrm{Jy}$. Given the limited high-frequency sampling ($\nu\gg\nu_\mathrm{a}$) in the early epochs, the electron energy index $p$ is poorly constrained and highly degenerate with the break-smoothness parameter. We therefore used epoch~4, which has the broadest frequency coverage, to constrain $p$. For this epoch we adopted a uniform prior of $p\in\left[2.0,4.0\right]$ and obtained $p=2.91\pm0.10$. We then fix $p=2.91$ in the SED modeling for all epochs.

\begin{figure}[h]
\centering
\includegraphics[width=0.95\columnwidth]{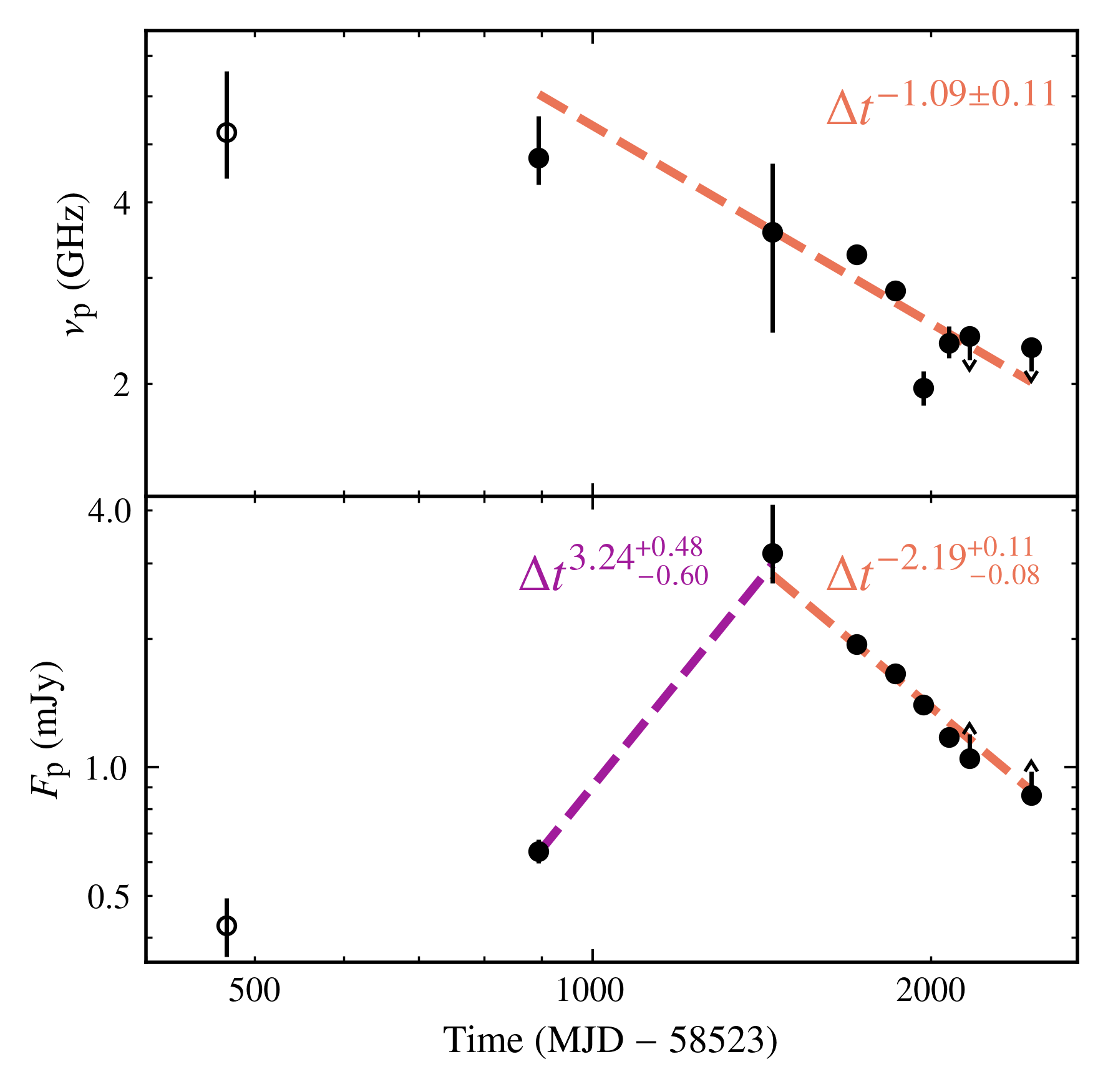}
\caption{Temporal evolution of the observed peak frequency, $\nu_{\mathrm{p}}$ (\textit{upper panel}) and peak flux density, $F_{\mathrm{p}}$ (\textit{lower panel}). Dashed lines indicate power-law fits to the data. The open symbol marks epoch~1, which is excluded from the fits. For the last two epochs, the SEDs do not constrain the spectral peak; the arrows show upper limits on $\nu_{\rm p}$ and lower limits
on $F_{\rm p}$.
\label{fig:fitting_results}}
\end{figure}

The multi-epoch radio spectra are well fit by the SSA model, with the resulting fits shown in Fig.~\ref{fig:sed_fitting}. From the posterior samples, we derived the observed peak frequency $\nu_\mathrm{p}$ and peak flux density $F_\mathrm{p}$ for each epoch. The spectral peak is not directly detected in the last two epochs. We therefore adopt conservative one-sided constraints, taking $F_{\rm p}>F_i$, where $F_i$ is the brightest observed flux density, and $\nu_{\rm p}<\nu_{i+1}$, where $\nu_{i+1}$ is the frequency of the next higher-frequency data point. To characterize the temporal evolution of $\nu_\mathrm{p}$ and $F_\mathrm{p}$, we fitted both curves with power-law models and found that epoch~1 significantly deviates from the fits to the other epochs. The best-fitting parameters are indicated in Fig.~\ref{fig:fitting_results}. While $\nu_\mathrm{p}$ exhibits a monotonic decline, $F_\mathrm{p}$ rises over the first $\sim1500$~d to a maximum of $\approx 3$~mJy before declining rapidly. The offset of epoch~1 suggests that the first radio detection may correspond to a different emission phase or even a distinct emission process. We therefore exclude it from the following analysis.

\subsection{Outflow-CNM:EquipartitionAnalysis}\label{subsec:equipartition}
We first consider an outflow-CNM model in which a single impulsively ejecta interacts with the CNM, driving a forward shock that powers the observed synchrotron radio emission. Using the equipartition formalism of \cite{BNT2013}, we infer the CNM properties from the measured spectral peak parameters ($\nu_\mathrm{p}$, $F_\mathrm{p}$) and the electron index $p$. The emission is assumed to arise from a thin spherical shell with radius $R$ and thickness $\Delta R=0.1R$, adopting an area filling factor $f_\mathrm{A}=1$ and a volume filling factor $f_\mathrm{V}\approx0.36$. Post‑shock microphysics is parameterized by the dimensionless energy fractions $\epsilon_\mathrm{e}$ and $\epsilon_B$, defined as the fractions of post‑shock internal energy deposited into non‑thermal electrons and magnetic fields, respectively. We adopt fiducial values $\epsilon_\mathrm{e}=\epsilon_B=0.1$ following \citet{Cendes2024}. Under these assumptions, the radius of the emitting region is:
\begin{equation}
\begin{aligned}
R \approx{} & \left(1 \times 10^{17}\,\mathrm{cm}\right)
\times \left(21.8 \times 525^{p-1}\right)^{\frac{1}{13+2p}}
\gamma_\mathrm{m}^{\frac{2-p}{13+2p}}
\\
&\times F_{\mathrm{p},\,\mathrm{mJy}}^{\frac{6+p}{13+2p}}
d_{\mathrm{L},28}^{\frac{2(p+6)}{13+2p}}
\nu_{\mathrm{p},10}^{-1}
(1+z)^{-\frac{19+3p}{13+2p}}
\\
&\times f_{\mathrm{A}}^{-\frac{5+p}{13+2p}}
f_{\mathrm{V}}^{-\frac{1}{13+2p}}
4^{\frac{1}{13+2p}}\Gamma^{\frac{p+8}{13+2p}}
\epsilon^{\frac{1}{17}},
\end{aligned}
\end{equation}
where $F_{\mathrm{p},\,\mathrm{mJy}}$ is the peak flux density in mJy, $d_{\mathrm{L},28}$ is the luminosity distance in units of $10^{28}$\,cm, $\nu_{\mathrm{p},10}$ is the observed peak frequency in units of 10\,GHz, $z$ is the redshift, and $\Gamma$ is the bulk Lorentz factor of the emitter. Following \citet{Cendes2024}, deviations from equipartition are captured by $\epsilon=(11/6)\epsilon_B/\epsilon_\mathrm{e}$, which modifies the radius by a factor of $\epsilon^{1/17}$. We also include the factor of $4^{1/(13+2p)}$ to account for the isotropic electron distribution in the non-relativistic case. The minimum electron Lorentz factor is $\gamma_\mathrm{m}=\mathrm{max}[\chi_\mathrm{e}(\Gamma-1),2]$, where $\chi_\mathrm{e}$ is defined as 
\begin{equation} 
\chi_\mathrm{e}=\frac{p-2}{p-1}\epsilon_\mathrm{e} \frac{m_\mathrm{p}}{m_\mathrm{e}}, 
\end{equation} 
with $m_\mathrm{p}$ and $m_\mathrm{e}$ denoting the proton and electron masses, respectively. 


We then begin with a Newtonian initial estimate by setting $\Gamma=1$ and $\gamma_\mathrm{m}=2$. A linear fit to $R(t)$ implies an approximately constant expansion velocity of $v=0.0124\pm0.0008c$, and yields an inferred launch time of $t_0=\,58803.5^{+79.6}_{-90.4}$. Given $t_0$, the shock velocity at each epoch is estimated as
\begin{equation}
v_\mathrm{s}=(1+z)\frac{R}{(t_\mathrm{obs}-t_0)}.
\end{equation}
At at all epochs, $v_\mathrm{s}/c\ll1$, and $v_\mathrm{s}$ is consistent with the mean $v$ within $3\sigma$, indicating no significant acceleration or deceleration. We therefore fix $\Gamma=1$ and $\gamma_\mathrm{m}=2$. With $R$ determined at each epoch, we infer the magnetic-field strength $B$, the Lorentz factor of electrons radiating at $\nu_\mathrm{a}$ ($\gamma_\mathrm{a}$), the number of radiating electrons $N_\mathrm{e}$, the energy in radiating electrons $E_\mathrm{e}$, and the magnetic energy $E_B$ as follows (see Eqs.~14--18 in \citealt{BNT2013}):

\begin{align}
\gamma_{\mathrm{a}} 
&\approx 525 \,
F_{\mathrm{p},\,\mathrm{mJy}}
d_{\mathrm{L},28}^{2}
\nu_{\mathrm{p},10}^{-2}
(1+z)^{-3}
f_\mathrm{A}^{-1}
R_{17}^{-2}
\Gamma, \\
N_{\mathrm{e}} 
&\approx (4 \times 10^{54})
F_{\mathrm{p},\,\mathrm{mJy}}^{3}
d_{\mathrm{L},28}^{6}
(1+z)^{-8} \notag \\
&\quad \times
f_{\mathrm{A}}^{-2}
R_{17}^{-4}
\nu_{\mathrm{p},10}^{-5}
\left(\frac{\gamma_{\mathrm{m}}}{\gamma_{\mathrm{a}}}\right)^{1-p}, \\
B 
&\approx 0.013\,
F_{\mathrm{p},\,\mathrm{mJy}}^{-2}
d_{\mathrm{L},28}^{-4}
(1+z)^{7}
\nu_{\mathrm{p},10}^{5}
f_{\mathrm{A}}^{2}
R_{17}^{4}
\Gamma^{-3}~\mathrm{G}, \\
E_\mathrm{e} 
&= N_\mathrm{e} m_\mathrm{e} c^2 \gamma_\mathrm{m} \Gamma, \\
E_\mathrm{B} 
&= \frac{(B\Gamma)^2}{8\pi} V.
\end{align}

\begin{figure}[h!]
\centering
    \includegraphics[width=0.95\columnwidth]{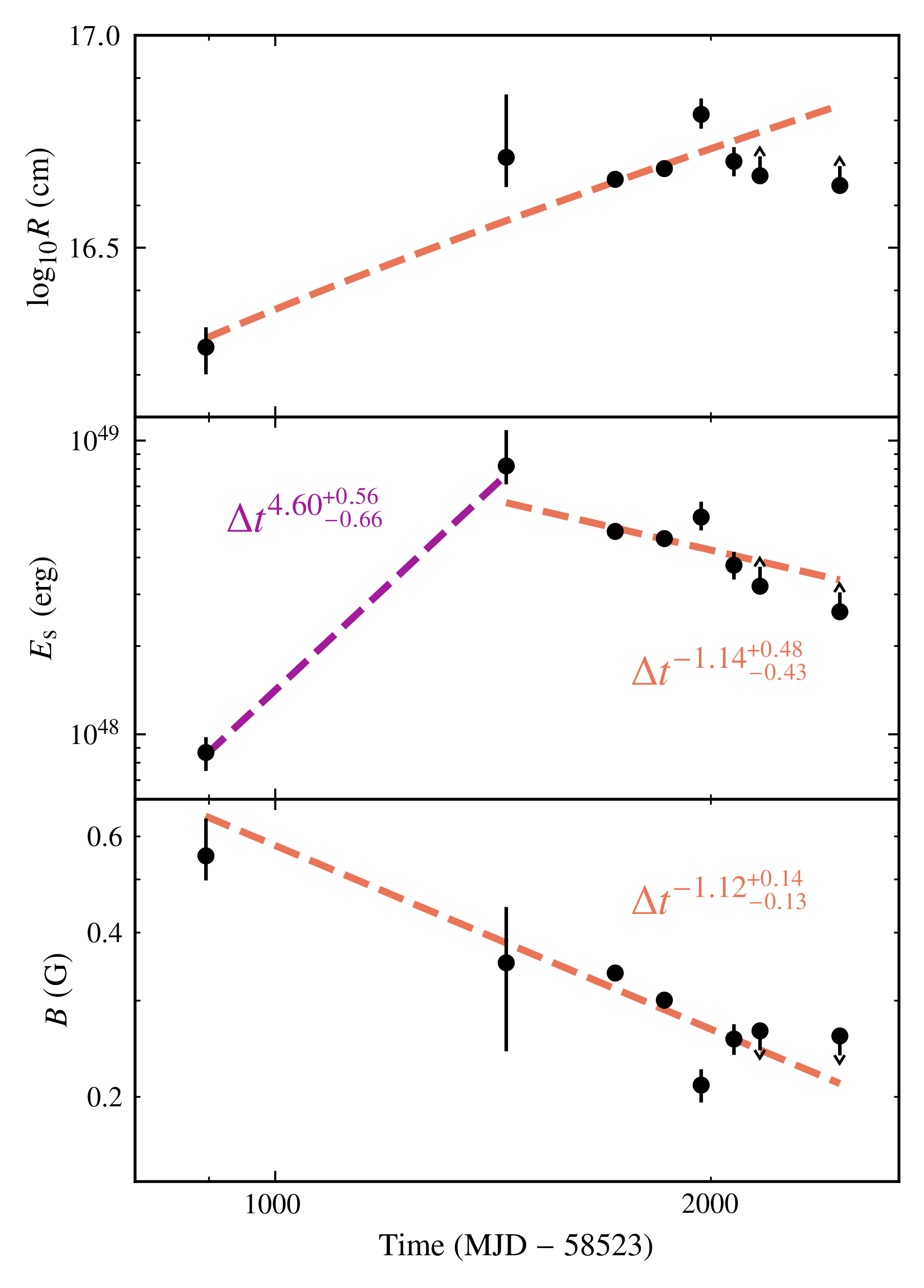}
    \caption{Time evolution of the emitting-region radius $R$, post-shock thermal energy $E_\mathrm{s}$, and magnetic-field strength $B$ inferred from the equipartition analysis. The lower and upper limits are indicated with arrows. In the top panel, the dashed line shows a linear fit of $R$ as a function of $t_\mathrm{obs}-t_0$, where $t_0$ is a free parameter representing the outflow launch time. In the other panels, power‑law fits are shown for $E_\mathrm{s}$ and $B$ as functions of $\Delta t\equiv t_\mathrm{obs}-t_\mathrm{d}$, with $t_\mathrm{d}=58523.23$ denoting the time of optical discovery.}
    \label{fig:physics_parameters}
\end{figure}

Here, $V=f_\mathrm{V}\pi R^3$ is the volume of the emitting region and $R_{17}\equiv R/10^{17}\,\mathrm{cm}$. In computing $N_\mathrm{e}$, we included a correction factor of $4(\gamma_\mathrm{m}/\gamma_a)^{1-p}$ to account for the isotropic number of radiating electrons in the Newtonian regime and for $\gamma_\mathrm{m}<\gamma<\gamma_\mathrm{a}$ (see \S\,4.2.1 of \citealt{BNT2013}). The post-shock thermal energy is then estimated as $E_\mathrm{s}=E_\mathrm{B}/\epsilon_\mathrm{B}$.The resulting temporal evolution of $R$, $E_{\rm s}$, and $B$ is shown in Fig.~\ref{fig:physics_parameters}. 

Since the thermal energy is approximately equal to post-shock kinetic energy, we could derive the ambient densities $n_{\mathrm{ext}}$ as:
\begin{equation}
\begin{aligned}
n_\mathrm{ext} = \frac{M_\mathrm{s}}{4Vm_\mathrm{p}}=\frac{2E_\mathrm{s}}{(\frac{3}{4}v_\mathrm{s})^2\times 4Vm_\mathrm{p}},
\end{aligned}
\label{eq:n_ext}
\end{equation}
where $M_\mathrm{s}$ is the shocked CNM mass, and the factors of 4 and $\frac{3}{4}$ account for the compression ratio and post-shock velocity under strong-shock conditions (Appendix B in \citealt{Mou_2026}). The derived physical parameters are listed in Table~\ref{tab:SED_equipartition}.

It is worth noting that our analysis adopts a different definition of the minimum energy and a different method for inferring the ambient density compared to some previous studies (e.g., \citealt{Cendes2021a, Cendes2024}). 
First, we employ the minimum-energy formalism in the \emph{two-component} form \citep{BNT2013}, defined as $E_\mathrm{eq,\,2}\equiv E_\mathrm{e}+E_\mathrm{B}$, which includes only the energy in radiating electrons and magnetic fields. This contrasts with the commonly used \emph{three-component} form, $E_\mathrm{eq,\,3}\equiv E_\mathrm{e}+E_\mathrm{B}+E_\mathrm{p}$, which additionally includes the energy of hot protons, $E_\mathrm{p}\approx E_\mathrm{e}/\epsilon_\mathrm{e}$. 
In most studies, the choice to define $E_\mathrm{p}\approx E_\mathrm{e}/\epsilon_\mathrm{e}$ means that $\epsilon_\mathrm{e}$ and $\epsilon_\mathrm{B}$ no longer represent the fractions of post‑shock thermal energy carried by electrons and magnetic fields, respectively (see Appendix B in \citealt{Mou_2026} for more details). We therefore adopt the \emph{two-component} form so that $E_\mathrm{e}$ and $E_\mathrm{B}$ remain consistent with the chosen $\epsilon_\mathrm{e}$ and $\epsilon_\mathrm{B}$.

Second, we infer the CNM density, $n_\mathrm{ext}$ (Eq.~\ref{eq:n_ext}), from the kinetic energy of the shocked CNM rather than from the non-thermal electron density, $n_\mathrm{nte}=N_\mathrm{e}/(4V)$, which is derived from the number of synchrotron-emitting electrons following a relativistic power-law distribution. 
The low expansion velocity inferred above, well below the critical value $\beta_{\rm DN}\sim0.2$ \citep{Rahaman2026}, indicates that the shocks are in the deep-Newtonian regime. In this regime, only a small fraction of the shocked electrons remain relativistic, such that $n_{\rm nte}$ may underestimate the true ambient density by factors of tens to hundreds \citep{matsumoto21}. Hereafter, we adopt $n_{\rm ext}$ as the estimate of the CNM density, while $n_{\rm nte}$ is also shown for comparison.

\begin{figure*}[htb!]
\centering
    \includegraphics[width=0.95\textwidth]{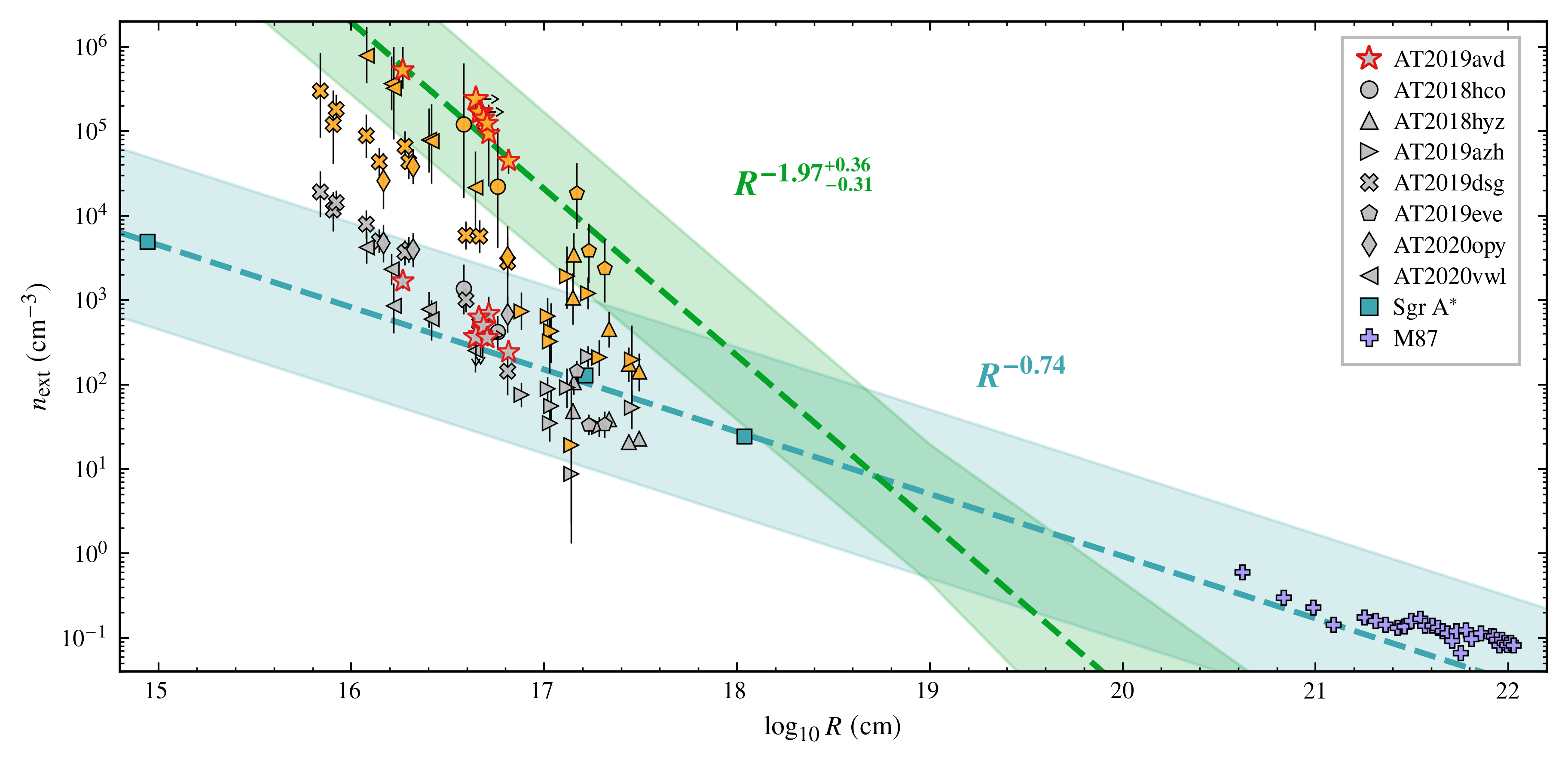}
\caption{The CNM density profiles of \avd and the comparison sample of radio TDEs. Yellow markers show ambient densities $n_{\rm ext}$, inferred from the post-shock kinetic energy using Eq.~\ref{eq:n_ext}, while gray markers show the non-thermal electron population $n_\mathrm{nte}$. The green dashed line shows the best-fitting power-law model for the $n_{\rm ext}$ profile of \avd, with the shaded region indicating the 16th--84th percentile posterior interval. For comparison, the density profiles of Sgr~A$^*$ \citep{Baganoff2003,Gillessen2019} and M87 \citep{Russell2015} are also shown. The cyan shaded region denotes $\pm1$\,dex around the best-fitting profile of Sgr~A$^*$. }
    \label{fig:density_profile}
\end{figure*}

To examine this difference across a broader sample, we selected TDEs from \citet{Cendes2024} with multi-epoch constraints on their synchrotron spectral peaks, yielding seven comparison sources, i.e., AT~2018hco \citep{Cendes2024}, AT~2018hyz \citep{Cendes2022}, AT~2019azh \citep{Goodwin2022}, AT~2019dsg \citep{Cendes2021a}, AT~2019eve \citep{Cendes2024}, AT~2020opy \citep{Goodwin2023a}, and AT~2020vwl \citep{Goodwin2023b}. We adopted the spectral parameters $(\nu_{\mathrm{p}}\,,F_{\mathrm{p}}\,,p)$ directly from these studies and recomputed both $n_{\rm ext}$ and $n_{\rm nte}$ using the same equipartition formalism, source geometry, and fiducial microphysical parameters as those adopted for \avd. The results are shown in Fig.~\ref{fig:density_profile}. Across the comparison sample, $n_{\rm ext}$ is systematically higher than $n_{\rm nte}$. 

\subsection{Outflow-cloud}\label{sec:Outflow-Cloud}
We then considered the outflow-cloud interaction model \citep{mou2021, mou2022}. In this model, the high-velocity outflow collides with clouds in the vicinity of the SMBH, driving bow shocks on the windward side of the clouds. The radio emission is generated by relativistic electrons accelerated at the bow shock front. The source exhibits an IR echo lagging the optical flare by several tens of days, indicating the presence of dust clouds at a distance of approximately 0.03\,pc from the SMBH \citep{Wang2023}. This provides the necessary conditions for the formation of bow shocks. 

We employ the ZEUS3D code \citep{clarke2010} to perform hydrodynamical simulations incorporating relativistic electrons, aiming to investigate whether the radio emission produced by the bow shock can reproduce the observed radio spectrum. 
The electron acceleration efficiency $\tilde\epsilon_\mathrm{e}$ is assumed to be the fraction of the energy flux that can be dissipated at the shock (i.e., the change in the kinetic energy flux across the shock) channeled into the relativistic electrons in the downstream. Relativistic electrons are injected into the simulation domain from the shock front grids, with their energy density set to be $e_{\rm cre}=5/3 \tilde\epsilon_\mathrm{e} e_d$, where $e_d$ is the thermal energy density in the post-shock gas \citep{mou2025b}, and $\tilde\epsilon_\mathrm{e}=0.04$. 
We assume magnetic fields are dynamically weak and fix a fiducial value of $\epsilon_\mathrm{B}= 0.1$. Even so, these fields remain important for relativistic electron cooling and synchrotron emission, and we include them in post-processing when computing the radio spectra \citep{Mou_2026}.
We modeled the cloud as a toroidal structure orbiting the SMBH, with a core distance of $d_\mathrm{c}$ and a cloud radius of $R_\mathrm{c}$. Given that the IR luminosity of the second flare exceeds that of the first, we speculate that a more powerful outflow was launched during the second outburst. Thus, the onset of the outflow is set to coincide with the second flare at MJD$=$59000. The outflow is injected isotropically from the inner boundary of $r_{\rm in} =3\times 10^{-3}$\,pc into the simulation domain, with the injection sustained for one year, and this duration is basically in line with the second flare's optical plateau stage. 

\begin{figure}[htb!]
\centering
\includegraphics[width=0.9\columnwidth]{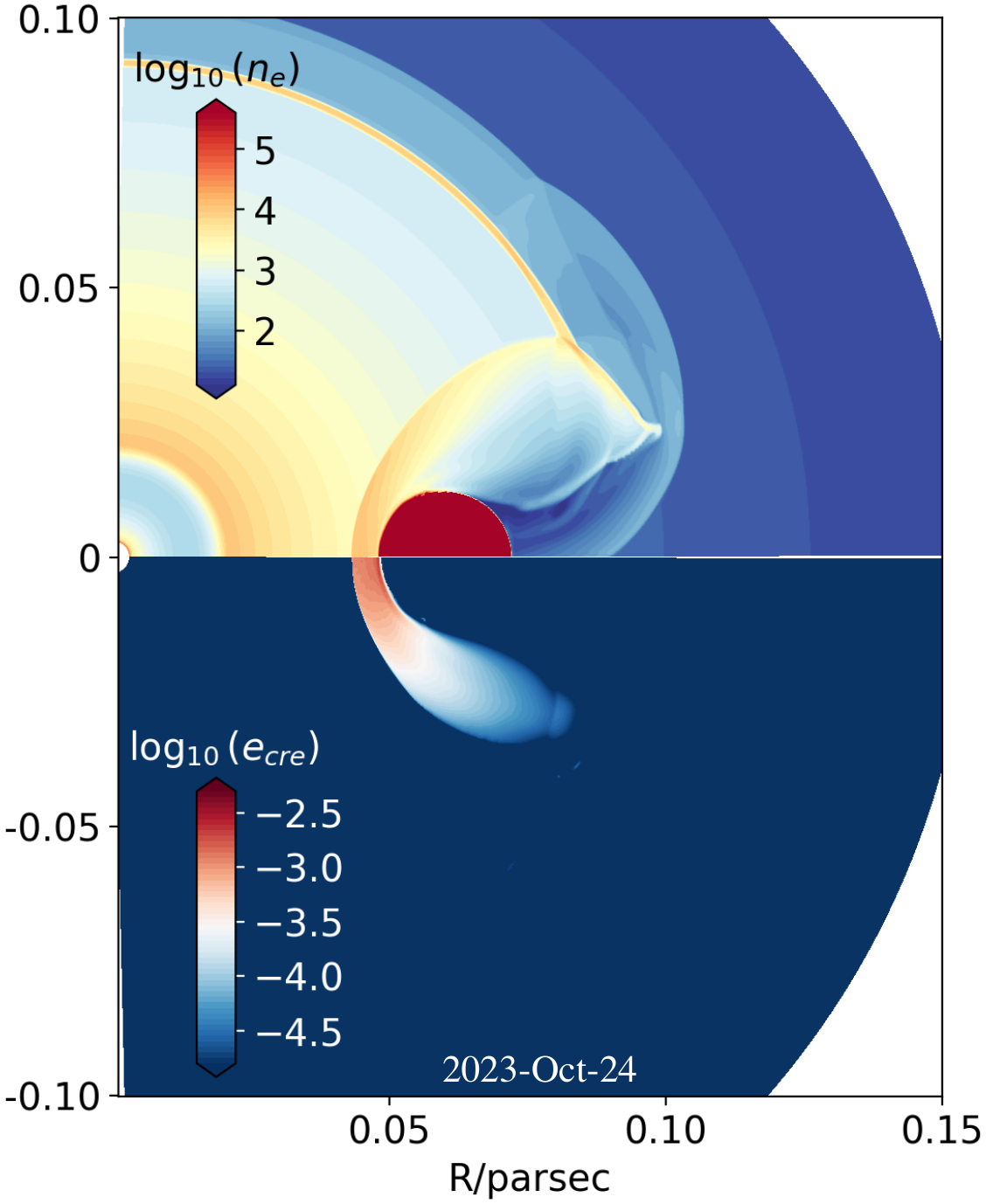}
\includegraphics[width=0.9\columnwidth]{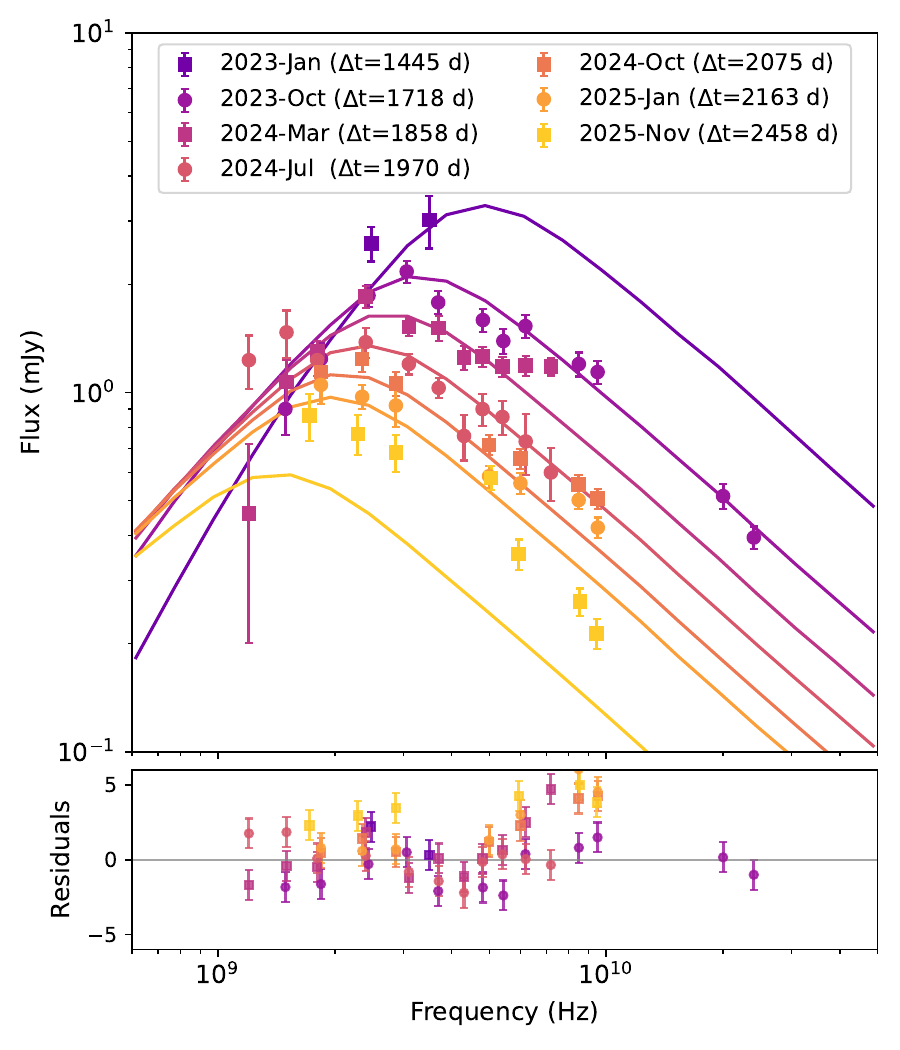}
 \caption{Numerical modeling the radio emission of \avd. \emph{Upper panel}: the thermal gas density $n_\mathrm{e}$ (in units of $\rm {cm^{-3}}$) and the energy density of relativistic electrons $e_{\rm cre}$ (in units of $\rm {erg~cm^{-3}}$) for the snapshot of 2023-Oct-24 ($\Delta t=1718$ d). \emph{Lower panel}: Comparison between the observed radio spectra and the synthetic spectra at the same epochs. The normalized residuals of (observation-model)/error are shown in the bottom window. } 
 \label{fig5}
\end{figure}

After testing multiple sets of model parameters, we identify a primary set that generally matches the radio observations (Fig.~\ref{fig5}): $d_\mathrm{c} = 0.06$\,pc and $R_\mathrm{c} = 0.012$\,pc. The outflow velocity decreases linearly from 0.1c to 0.025c, and the mass outflow rate is $\dot{M}_{\rm out}=0.27\,\msunyr$ for the first 30\,days, after which it declines overtime linearly from 0.17 to 0.10\,$\msunyr$. This setup yields a total outflow mass of 0.145\,$M_{\odot}$, and a cumulative kinetic energy of $6.5\times 10^{50}$\,erg. 

\section{Discussion}\label{sec:disscussion}
We analyze more than five years of radio observations of \avd and derive the physical parameters under both forward-shock and bow-shock scenarios. We discuss possible outflow origins from the outflow-CNM equipartition analysis and the associated tensions in both the outflow-CNM and outflow-cloud scenarios.

\subsection{Outflow-CNM scenario: outflow origin}
These two well‑separated optical flares, with X-ray brightening coincident with the second flare of \avd have motivated a ``two‑phase'' interpretation, in which the first flare is powered by stream self‑intersection (circularization) and the second by delayed accretion through a nascent disk \citep{Chen2022,Wang2023}. Moreover, \citet{Chen2022} estimated the onset of the accretion-dominated phase to be MJD~$58839_{-4}^{+3}$ from multi-wavelength modeling. Our equipartition analysis yields an outflow launch time of $t_0=58803.5^{+79.6}_{-90.4}$, consistent within the uncertainties with the inferred onset of the accretion-dominated phase. Within this context, we discuss the potential physical origins of the observed outflow.

\subsubsection{On/off-axis relativistic jets} \label{subsec:jet}


Both the low outflow velocity ($v\sim10^{-2}c$) and the sub-luminous radio emission ($L<2\times10^{38}$\,erg/s) of \avd argue against an on-axis relativistic jet scenario. We further consider the possibility of an off-axis jet using the generalized equipartition formalism for arbitrary viewing angles \citep{Matsumoto2023}. 

Within this framework, the minimum-energy solutions separate into Newtonian/on-axis and relativistic/off-axis branches, with a continuous connection between the two requiring the apparent Newtonian velocity parameter to exceed a critical value ($\beta_{\rm eq,N}\gtrsim0.44$; \citealt{Beniamini2023}). For AT~2019avd, we find $\beta_{\rm eq,N}\sim0.01$ at all epochs, well below this criterion, indicating that an off-axis relativistic solution cannot be smoothly connected to the observed Newtonian evolution. Combined with the observed late-time decline in both peak flux ($F_{\rm p}$) and the post-shock thermal energy ($E_{\rm s}$), an off-axis jet would therefore require a rapid deceleration and an abrupt transition to the Newtonian phase between epochs~2 and~3 ($t\sim1000$--$1500$\,days), for which we find no supporting evidence.

To quantitatively test the off-axis jet scenario, we performed forward modeling with VegasAfterglow \citep{Zhang2018, vegasafterglow}. We adopted a uniform top-hat jet characterized by an isotropic-equivalent kinetic energy $E_\mathrm{iso}$, an initial Lorentz factor $\Gamma_0$, a half-opening angle $\theta_\mathrm{c}$, and a viewing angle $\theta_\mathrm{v}$. The jet propagates into a power-law CNM density profile parameterized by a normalization $A_\star$, and an index, $k_{\rm m}$. The microphysical parameters $\epsilon_{\rm e}$, $\epsilon_{\rm B}$, and $\xi_{\rm e}$, where $\xi_{\rm e}$ denotes the fraction of electrons accelerated into the non-thermal distribution, were assumed to remain constant throughout the modeled period.

\begin{figure}
\centering
\includegraphics[width=0.4\textwidth]{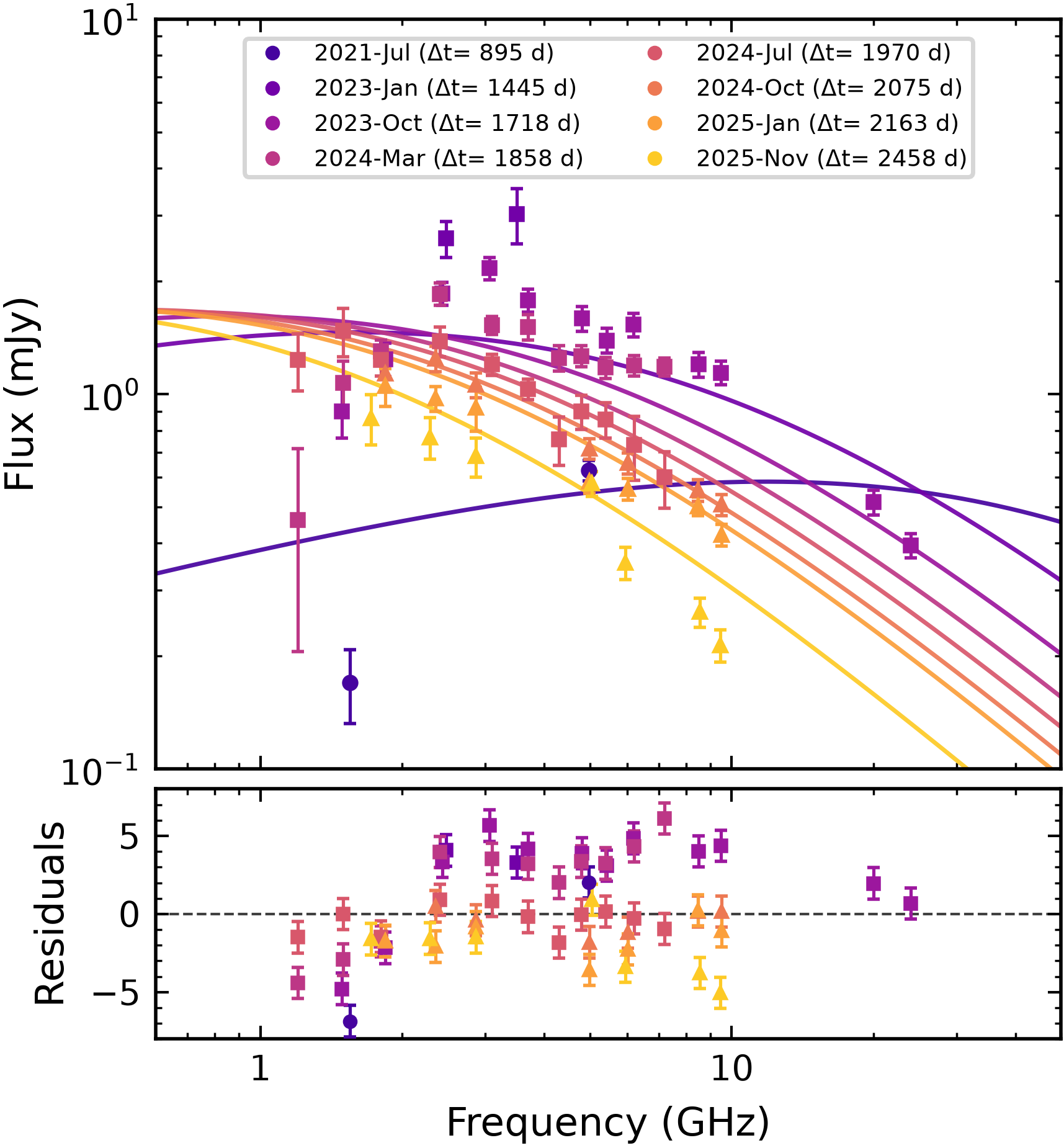}
\includegraphics[width=0.4\textwidth]{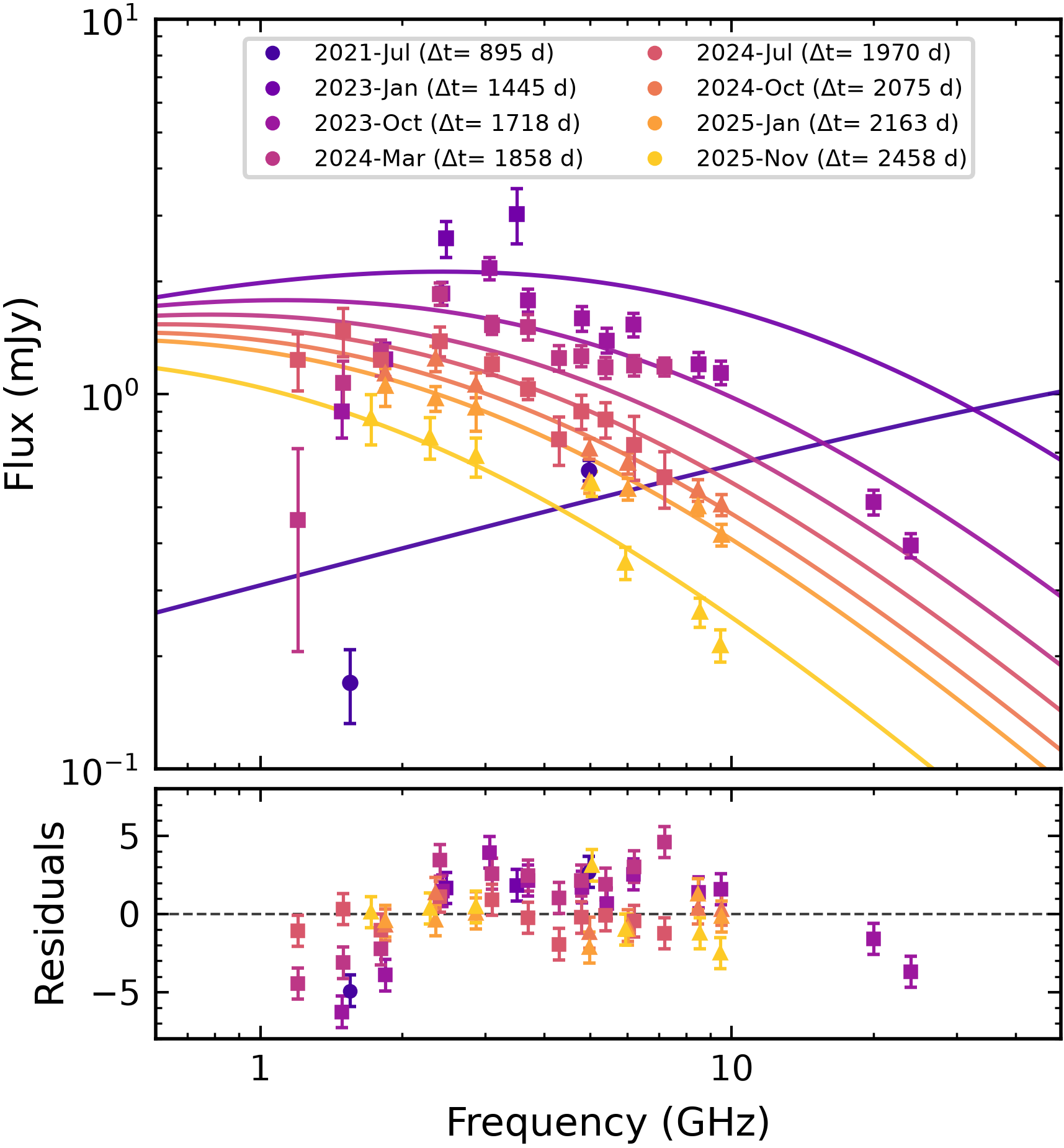}
\caption{Radio SED fits of \avd\ using top-hat jet models computed with \texttt{VegasAfterglow}. The upper and lower panels correspond to fixed launch times of $t_{\rm 0}=58803.5$ and $t_{\rm 0}^{\prime}=59335$, respectively. In each panel, the upper subpanel shows the observed flux densities and the corresponding best-fitting model SEDs, while the lower subpanel shows the residuals, defined as $(F_{\nu,\,\rm obs}-F_{\nu,\,\rm mod})/\sigma_{F_\nu,\,\rm obs}$.}
\label{fig:vegas_fit}
\end{figure}

We fitted the radio flux-density measurements with \texttt{emcee} \citep{emcee}, considered two possible launch times: $t_{\rm 0}=58803.5$, inferred from the equipartition radius evolution, and $t_{\rm 0}^{\prime}=59335$, motivated by the onset of the late hard X-ray state reported by \citet{Wang2024}. The adopted priors and resulting posterior constraints for both fits are summarized in Table~\ref{tab:vegas_priors}.

The models with $t_{\rm 0}=58803.5$ and $t_{\rm 0}^{\prime}=59335$ yield $\chi^2_\nu=10.42$ and $5.81$, respectively. The later-launch model therefore provides a better fit, although both fits remain statistically unacceptable. As shown in Fig.~\ref{fig:vegas_fit}. the models systematically overpredict the low-frequency flux densities at $\sim$ 1--2\,GHz and fail to reproduce the observed spectral shape around the turnover. 

We therefore disfavor the specific off-axis jet model tested here. More complex configurations, including an angularly structured jet, a structured or non-monotonic CNM, evolving microphysical parameters, or additional low-frequency absorption, may provide a better fit. However, the current data are insufficient to constrain these additional degrees of freedom.

\subsubsection{Unbound debris} 

In canonical TDEs, approximately half of the disrupted stellar mass is expected to become unbound and ejected right after the disruption \citep{Rees1988}. 

Simulations show that the maximum speed of the escaping material increases with the penetration factor, $R_\mathrm{T}/R_\mathrm{p}$, where $R_{\rm T}$ and $R_{\rm p}$ represent the disruption and the pericenter radii, respectively \citep{Yalinewich2019}. For a marginal encounter ($R_\mathrm{T}/R_\mathrm{p} \simeq 1$), the fastest debris reaches $v_{\max}\lesssim 0.017c$, whereas even a deeply penetrating disruption ($R_\mathrm{T}/R_\mathrm{p} \simeq 7$) produces $v_{\max} \lesssim 0.07c$. For \avd, optical light-curve modeling favours a partial disruption with penetration factor $R_\mathrm{T}/R_\mathrm{p}\simeq0.56$ \citep{Chen2022}, for which the unbound debris is expected to move more slowly and occupy a small solid angle. If we instead adopt a small emitting-area filling factor ($f_\mathrm{A}= f_\mathrm{V}=0.008$, based on the $R_\mathrm{T}/R_\mathrm{p}\simeq1$ case in  \citealt{Yalinewich2019}) in the equipartition analysis, we infer a mean expansion speed is $v=0.116\pm0.007c$, far above expectations for unbound debris. The corresponding launch time, $58803.4^{+79.9}_{-89.7}$, is approximately 280\,days after the optical discovery, whereas unbound debris would be produced essentially immediately after the disruption. We therefore disfavor an unbound debris origin for the outflow responsible for the AT~2019avd radio flare.

\subsubsection{CIO or accretion-driven outflow} 
Stream self-intersection is capable of both producing thermal optical/UV emission and driving a quasi-spherical outflow. Subsequently, radio emission is generated as the outflow interacts with the CNM, giving rise to a SSA spectrum whose peak time and luminosity depend on the launch conditions and ambient density \citep{Lu2020}. In \avd, the inferred outflow launch time, $t_0$, coincides with the onset of the delayed accretion phase. This temporal association therefore disfavors a CIO launched during the initial optical flare as the dominant origin of the radio-emitting outflow.
 
Simulations predict powerful, wide‑angle winds during super‑Eddington accretion \citep{Dai2018}, and a growing sample of optical TDEs exhibits delayed radio peaks (hundreds-thousands of days) plausibly launched by delayed accretion or accretion state transition \citep{Cendes2024,Alexander2026}. In \avd, the agreement between the inferred outflow launch time, $t_0$, and the onset of the accretion-dominated phase suggests that the radio-emitting outflow was launched as the accretion flow formed. We therefore identify a delayed, accretion-driven outflow as the most plausible origin of the radio emission in the outflow-CNM interaction scenario.

\subsection{Potential tensions with the outflow-CNM and outflow-cloud scenarios}

\subsubsection{The decrease of the post-shock energy}
The post-shock energy exhibits a decline after epoch~3 (middle panel of Fig.~\ref{fig:physics_parameters}). To assess whether the forward shock has entered the momentum-conserving snowplow phase, we estimate the cooling timescale as follows.

The forward shock velocity of \avd\ is estimated to be $\sim 0.01c$, corresponding to a post-shock CNM temperature of $T_\mathrm{s} = \frac{3}{16} \frac{\mu m_\mathrm{p}}{k_\mathrm{B}} v_s^2 \approx 1.2 \times 10^8 \text{ K} \left( \frac{v_\mathrm{s}}{0.01c} \right)^2$. 
Given the thermal energy density $e=\frac{3}{2}(n_\mathrm{e}+n_\mathrm{i}) k_\mathrm{B} T_\mathrm{s}$, where $n_\mathrm{e}$ and $n_\mathrm{i}$ are number densities of thermal electrons and ions, the cooling timescale of post-shock CNM is $t_{\rm cool}=e/C$.  
The cooling rate $C$ includes contributions from bremsstrahlung, line cooling, inverse Compton scattering and synchrotron: 
\begin{equation}
C= \Lambda n_\mathrm{i} n_\mathrm{e} + \frac{4}{3}n_\mathrm{e} \sigma_T c (u_{\rm rad} + u_{\rm mag}) (v_{\rm th}/c)^2, 
\end{equation}
where $\Lambda$ is the cooling function \citep{Sutherland1993}, $u_{\rm rad}=L/4\pi R^2 c$ is the radiation energy density, $u_{\rm mag}=B^2/8\pi$, and $v_{\rm th}$ is the thermal velocity. 
Adopting representative values at $\Delta t\sim 1500$d, $\Lambda \simeq {3-6}\times 10^{-23}~ {\rm erg~s^{-1}\,cm^3}$ (for a temperature of $10^{8-9}$ K), $L < 10^{42}\,{\rm erg~s^{-1}}$, and $B=0.4$\,G, we find that bremsstrahlung dominates the cooling, giving $t_{\rm cool} > 100$\,years. 

The long cooling timescale indicates that the forward shock has not yet entered the momentum-conserving snowplow phase, but remains in either the free-expansion phase or the Sedov-Taylor phase, depending on the ejecta mass. In this regime, the post-shock energy is expected to increase or remain approximately constant, rather than decline with time. However, the inferred decline assumes time-independent shock microphysics. If $\epsilon_{\rm e}$ decreases with time, the true shock energy could instead remain constant or even increase, a possibility that we cannot completely rule out.

\subsubsection{The rapid rise and decline radio light curve}

\begin{figure}
\centering
    \includegraphics[width=0.95\columnwidth]{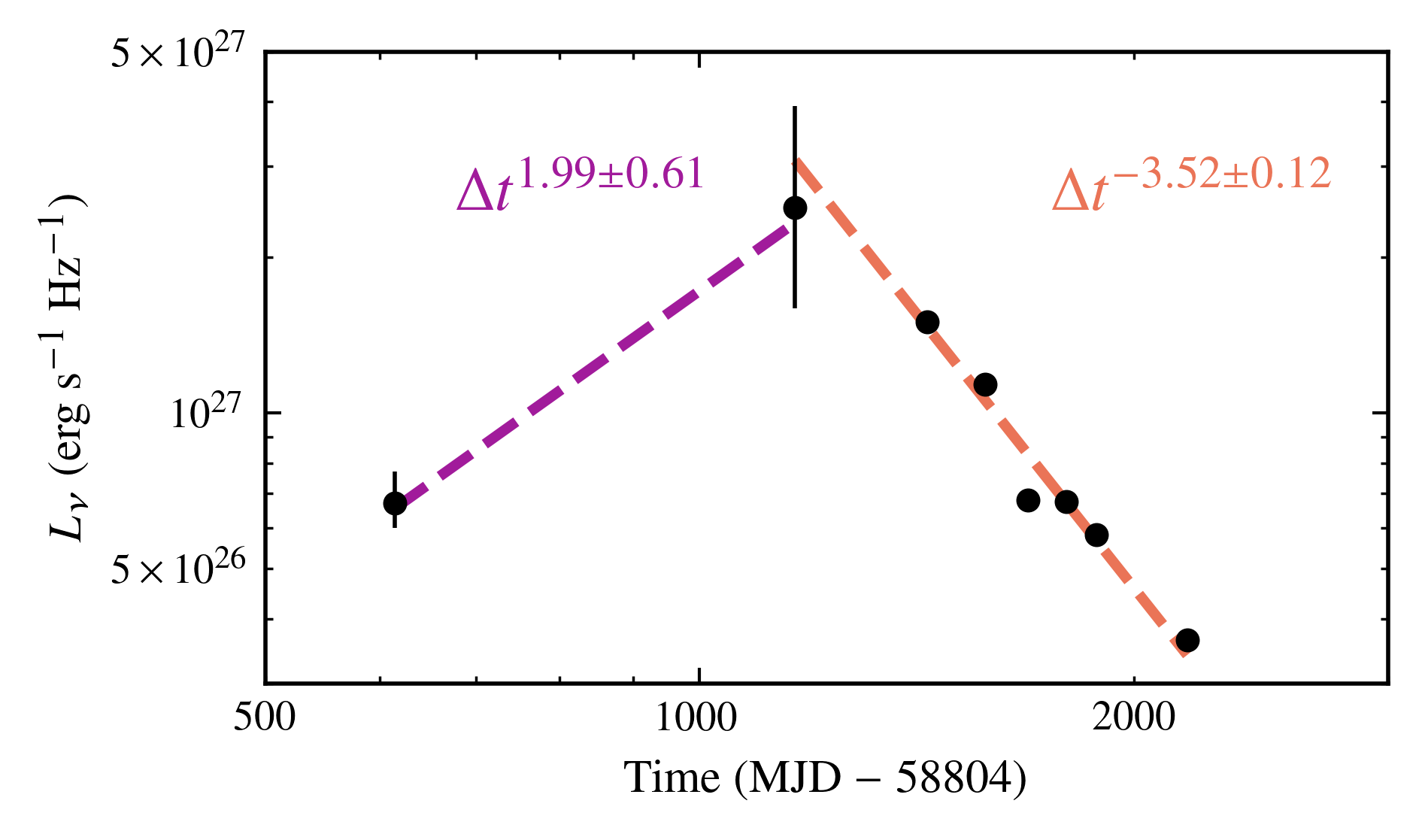}
    \caption{The $9\,\mathrm{GHz}$ light curve inferred from the SED model (Sect.~\ref{subsec:spec_fit}). The dashed lines show  power-law fits to the different temporal phases (two rising segments followed by a decay).}
    \label{fig:light_curve}
\end{figure}

Rapid temporal evolution of the radio luminosity has been proposed as an additional challenge to the outflow-CNM scenario, which predicts a steepest rise of only $L \propto t^3$ (e.g., \citealt{Berger2012,Zhuang2025}). To quantify the radio light-curve evolution of \avd, we reconstructed the 9\,GHz monochromatic luminosity using the posterior samples from the multi-epoch SSA spectral fits (Sect.~\ref{subsec:spec_fit}), and modeled its temporal evolution with a broken power law, $L_\nu\propto (t-t_0)^\alpha$, where $t_0=$58803.5. The best-fit model yields a rapid rise with $\alpha_r=1.99\pm0.61$, followed by a steep decay with $\alpha_d = -3.52\pm0.12$. Within the outflow-CNM framework, \cite{Mou_2026} derives a relationship between the temporal evolution index, $\alpha$, and the CNM density profile, $k$ (see their Sect.~5.3). Adopting $p=2.91$, we infer $k_r=-0.51\pm0.31$ during the rising phase and $k_d=-3.33\pm0.06$ during the declining phase of the \avd light curve. This evolution implies a relatively flat density profile at smaller radii, followed by a much steeper decline at larger radii, opposite to the trend expected in steady-state stellar-wind-fed CNM models, in which the density profile generally flattens toward larger radii \citep{Generozov2017}.
By contrast, the outflow-cloud interaction scenario naturally predicts both rapid brightening and fading phases \citep{mou2021,mou2025b}, motivating the exploration of this alternative interpretation in Sect.~\ref{sec:Outflow-Cloud}.

\subsubsection{The relatively high density profile}
For comparison with TDEs, we also include the density profile for the two of the best-studied galactic nuclei, Sgr~$A^{*}$ and M87, in Fig.~\ref{fig:density_profile}. 
At the characteristic CNM radii of $R\sim10^{16-17}\,\mathrm{cm}$, the comparison TDEs generally exhibit steeper radial density gradients than those inferred for Sgr~A$^{*}$ and M87, while also occupying a substantially higher-density regime. To quantify this difference, we fit the density profiles of \avd and Sgr~$A^{*}$ with a single power law, $n_\mathrm{ext}\propto R^{k}$, obtaining $k=-1.97^{+0.36}_{-0.31}$ for \avd, and $k\sim-0.74$ for Sgr~$A^{*}$.

A steep density profile is not unique to \avd. In our comparison sample, the best-fitting density indices span a broad range, from $k\sim-0.8$ to $-4.8$, with four sources clustering around $k\sim-2$ to $-3$.
This diversity suggests that steep radial density gradients are relatively common among radio TDEs. However, systematic uncertainties nevertheless remain. Although $n_{\rm ext}$ is a more physically meaningful quantity than $n_{\rm nte}$ in the deep-Newtonian regime, its absolute normalization still depends on the assumed geometry, filling factors, and microphysical parameters. It therefore remains unclear whether the steep CNM profile inferred for TDEs are intrinsic or partly driven by methodological biases.



\subsubsection{The late-time SED deviation from the outflow-cloud scenario}
In the outflow-cloud model, the interaction between the outflow and a toroidal cloud located in the equatorial plane drives a bow shock on the inner (windward) face of the cloud. This shock efficiently accelerates electrons, leading to a substantial population of relativistic particles (Fig.~\ref{fig5}, see Fig.~\ref{figA2} for the time evolution). 

We compute synthetic synchrotron spectra in the polar direction using radiative transfer (see Appendix~\ref{sec:simu} of this paper and Sect.~4 in \citealt{Mou_2026} for more details). The resulting synthetic spectra are generally consistent with the multi-epoch radio data, although the discrepancy is largest for the final epoch.

The late-time deviations may arise from an additional contribution of the forward shock, which becomes increasingly important as the mass of the swept-up CNM approaches the ejecta mass. Alternatively, the outflow may have encountered an additional cloud at larger radii that is not included in our simulations, with the associated bow shock contributing extra radio emission. In the former scenario, the radio luminosity may not decline in future observations and could even increase.
 
We emphasize that the simulation parameters --including those describing the cloud and the outflow-- are not uniquely constrained, due to limited observational constraints and uncertainties in the microphysics (such as $\epsilon_\mathrm{e}$ and $\epsilon_B$). 
Our tests indicate that a cloud radius $R_\mathrm{c}$ in the range of 0.01--0.02\,pc can reproduce the observed spectra. As noted in \cite{mou2025b}, in the bow-shock scenario the characteristic radius $R$ derived from equipartition effectively traces the cloud size, rather than the forward-shock radius as in the outflow-CNM scenario.

Moreover, our tests show that smaller cloud radii lead to higher opacity, shifting the synchrotron self-absorption peak to higher frequencies. On the other hand, the peak frequency becomes relatively insensitive to further increase in cloud radius beyond a certain scale (Appendix~\ref{sec:simu}). Consequently, the upper limit of the cloud size remains uncertain.


\section{Conclusion} \label{sec:conclusion}
We presented a detailed study of the radio flare from \avd spanning more than five years. We modeled each epoch with a synchrotron self-absorbed spectrum and used the inferred spectral parameters to test outflow-CNM and outflow-cloud interaction scenarios. Our main conclusions are as follows:

\begin{enumerate}
    \item The multi-epoch radio spectra are well described by a single SSA component at each epoch. 
    The first radio detection deviates from the later evolution in both $\nu_{\rm p}$ and $F_{\rm p}$, which implies a different origin at early times. 
    
    \item In the outflow-CNM framework, our equipartition analysis suggests a non-relativistic expansion with $v\simeq 0.01c$ and a launch time of $\mathrm{MJD}\,58803.5^{+79.6}_{-90.4}$. The inferred kinematics and energetics disfavor both on-axis and off-axis relativistic jet scenarios, as well as an origin in unbound debris. The launch time instead favors an accretion-driven outflow associated with the delayed accretion phase.

    \item The equipartition results reveal several tensions with a simple single-blob outflow--CNM interaction scenario, but do not rule out this class of models in general. More complex scenarios involving evolving microphysical properties or a structured CNM remain viable.
    
    \item The outflow-cloud interaction scenario can reproduce most of the observed radio spectra. The presence of the required dense material is supported by the IR brightening coincident with the optical flares. The final epoch exhibits the largest deviation from the bow-shock model, indicating the emergence of an additional component at late times. This component may correspond to a forward shock that becomes increasingly important as the swept-up CNM mass grows, or to a second interaction between the outflow and another dense cloud.

\end{enumerate}

Overall, \avd supports a picture in which bow shocks can dominate the radio flare of some TDE candidates, while a forward shock may contribute increasingly at late times. However, the physical origin of the latest radio spectrum is unclear. Continued radio monitoring, can test the origin of the late-time emission. VLBI imaging may provide constraints on the source size and centroid motion predicted by off-axis jet models. Moreover, a larger sample is required to examine different implementations of the equipartition analysis.

\section*{Acknowledgements} 
We thank the anonymous referee for the constructive comments and suggestions that have improved and reshaped the manuscript. The authors thank the participants of the TDE FORUM (Full-process Orbital to Radiative Unified Modeling) online seminar series for their inspiring discussions. 
This research was supported by the National Natural Science Foundation of China (NSFC) under grant No. 12588202; by the New Cornerstone Science Foundation through the New Cornerstone Investigator Program and the XPLORER PRIZE; by the Strategic Priority Program of the Chinese Academy of Sciences under grant No. XDB0550203. 
G.M. was supported by the NSFC (No. 12473013). 
S.d.P. acknowledges support from ERC Advanced Grant 789410.

\software{
Python \citep{vanRossum2009Python},
Astropy \citep{astropy:2013,astropy:2018,astropy:2022},
NumPy \citep{harris2020array},
SciPy \citep{virtanen2020scipy},
pandas \citep{mckinney-proc-scipy-2010, reback2020pandas},
emcee \citep{emcee},
dynesty \citep{2020MNRAS.493.3132S,koposov_dynesty_zenodo},
bilby \citep{bilby_paper,bilby_doi},
Matplotlib \citep{Hunter:2007},
CASA \citep{CASA},
VegasAfterglow \citep{Zhang2018,vegasafterglow},
VLASS Scripts \citep{VLASS_Scripts}
}

\bibliography{cite}{}
\bibliographystyle{aasjournal}

\appendix
\restartappendixnumbering
\section{Radio observations}
We summarize the radio observations of \avd\ in Table~\ref{tab:radio_obs}.
\begingroup
\startlongtable
\begin{deluxetable*}{lccccc}
\tablecaption{Radio observations of \avd\label{tab:radio_obs}}
\tabletypesize{\small}
\tablewidth{1pt}
\renewcommand{\arraystretch}{0.92}
\tablehead{
\colhead{Date} & \colhead{Instrument} & \colhead{Proposal ID} &
\colhead{Frequency} & \colhead{\shortstack{Flux density}} & \colhead{Error} \\
\colhead{(UTC)} & \colhead{} & \colhead{} &
\colhead{(GHz)} & \colhead{(mJy)} & \colhead{(mJy)}
}
\startdata
2020-05-25$^{a}$ & VLA  & 20A-514 & 9.0  & 0.37   & 0.02 \\
                 &      &         & 11.1 & 0.27   & 0.02 \\
2020-08-09$^{a}$ & VLA  & VLASS 2 & 3.0  & $<$0.48 & -    \\
\tableline
2021-06-22$^{a}$ & VLBA & BW142   & 1.6  & 0.17   & 0.04 \\
2021-08-22$^{a}$ & VLBA & BW142   & 5.0  & 0.63   & 0.04 \\
\tableline
2023-01-24        & VLA & VLASS 3 & 2.5  & 2.60   & 0.29 \\
                  &     &         & 3.5  & 3.02   & 0.51 \\
\tableline
2023-10-10        & VLA & 23B-301 & 1.2  & $<$0.90 & -    \\
                  &     &         & 1.5  & 0.90   & 0.14 \\
                  &     &         & 1.8  & 1.24   & 0.13 \\
                  &     &         & 2.4  & 1.86   & 0.13 \\
                  &     &         & 3.1  & 2.17   & 0.15 \\
                  &     &         & 3.7  & 1.78   & 0.13 \\
                  &     &         & 4.8  & 1.59   & 0.12 \\
                  &     &         & 5.4  & 1.39   & 0.11 \\
                  &     &         & 6.2  & 1.53   & 0.11 \\
                  &     &         & 8.5  & 1.20   & 0.09 \\
                  &     &         & 9.5  & 1.14   & 0.08 \\
2023-11-07        & VLA & 23B-301 & 20.0 & 0.52   & 0.04 \\
                  &     &         & 24.0 & 0.40   & 0.03 \\
\tableline
2024-03-12        & VLA & 24A-434 & 1.2  & 0.46   & 0.26 \\
                  &     &         & 1.5  & 1.07   & 0.16 \\
                  &     &         & 1.8  & 1.30   & 0.09 \\
                  &     &         & 2.4  & 1.85   & 0.13 \\
                  &     &         & 3.1  & 1.53   & 0.09 \\
                  &     &         & 3.7  & 1.51   & 0.12 \\
                  &     &         & 4.3  & 1.25   & 0.10 \\
                  &     &         & 4.8  & 1.26   & 0.08 \\
                  &     &         & 5.4  & 1.18   & 0.08 \\
                  &     &         & 6.2  & 1.19   & 0.08 \\
                  &     &         & 7.2  & 1.18   & 0.07 \\
\tableline
2024-07-01        & VLA & 24A-434 & 1.2  & 1.23   & 0.21 \\
                  &     &         & 1.5  & 1.47   & 0.22 \\
                  &     &         & 1.8  & 1.23   & 0.12 \\
                  &     &         & 2.4  & 1.38   & 0.13 \\
                  &     &         & 3.1  & 1.20   & 0.08 \\
                  &     &         & 3.7  & 1.03   & 0.07 \\
                  &     &         & 4.3  & 0.76   & 0.11 \\
                  &     &         & 4.8  & 0.90   & 0.09 \\
                  &     &         & 5.4  & 0.86   & 0.09 \\
                  &     &         & 6.2  & 0.73   & 0.14 \\
                  &     &         & 7.2  & 0.60   & 0.10 \\
\tableline
2024-10-12        & ATCA & C3662  & 5.0  & 0.72   & 0.05 \\
                  &      &        & 6.0  & 0.65   & 0.04 \\
                  &      &        & 8.5  & 0.55   & 0.04 \\
                  &      &        & 9.5  & 0.51   & 0.03 \\
2024-10-16        & ATCA & C3615  & 1.8  & 1.14   & 0.11 \\
                  &      &        & 2.4  & 1.24   & 0.10 \\
                  &      &        & 2.9  & 1.06   & 0.08 \\
\tableline
2025-01-11        & ATCA & C3662  & 1.8  & 1.05   & 0.12 \\
                  &      &        & 2.4  & 0.97   & 0.08 \\
                  &      &        & 2.9  & 0.92   & 0.12 \\
                  &      &        & 5.0  & 0.59   & 0.04 \\
                  &      &        & 6.0  & 0.56   & 0.04 \\
                  &      &        & 8.5  & 0.50   & 0.03 \\
                  &      &        & 9.5  & 0.42   & 0.03 \\
\tableline
2025-10-22        & ATCA & C3755  & 1.7  & 0.86   & 0.13 \\
                  &      &        & 2.3  & 0.77   & 0.10 \\
                  &      &        & 2.9  & 0.68   & 0.08 \\
2025-11-11        & ATCA & C3755  & 5.1  & 0.58   & 0.04 \\
                  &      &        & 5.9  & 0.36   & 0.04 \\
                  &      &        & 8.6  & 0.26   & 0.02 \\
                  &      &        & 9.4  & 0.21   & 0.02 \\
\enddata
\tablecomments{Upper limits are given at the $3\sigma$ level. The symbol $a$ indicates flux densities taken from \citet{Wang2023}.}
\end{deluxetable*}
\endgroup

\section{Spectral Fitting Parameters and equipartition analysis results}
We summarize the radio SED fitting results (Sect.~\ref{subsec:spec_fit}) and the corresponding equipartition-derived physical parameters (Sect.~\ref{subsec:equipartition}) for each observing epoch in Table~\ref{tab:SED_equipartition}.

\begingroup
\startlongtable
\begin{deluxetable*}{lccccccc}
\tablecaption{Radio SED parameters and equipartition results.\label{tab:SED_equipartition}}
\tabletypesize{\small}
\tablehead{
\colhead{$\Delta t$} & \colhead{$\nu_{\rm p}$} & \colhead{$F_{\rm p}$} &
\colhead{$\log_{10} R$} & \colhead{$\log_{10} N_{\rm e}$} &
\colhead{$\log_{10} B$} & \colhead{$\log_{10} E_{\rm s}$} &
\colhead{$\log_{10} n_{\rm ext}$} \\
\colhead{(days)} & \colhead{(GHz)} & \colhead{(mJy)} &
\colhead{(cm)} & \colhead{} &
\colhead{(G)} & \colhead{(erg)} &
\colhead{(cm$^{-3}$)}
}
\startdata
895  & $4.74^{+0.82}_{-0.46}$ & $0.64^{+0.04}_{-0.04}$ & $16.27^{+0.05}_{-0.06}$ & $52.68^{+0.05}_{-0.06}$ & $-0.26^{+0.07}_{-0.05}$ & $47.94^{+0.05}_{-0.06}$ & $5.72^{+0.28}_{-0.22}$ \\
1445 & $3.57^{+1.07}_{-1.13}$ & $3.17^{+0.94}_{-0.47}$ & $16.71^{+0.15}_{-0.07}$ & $53.66^{+0.12}_{-0.06}$ & $-0.45^{+0.10}_{-0.16}$ & $48.91^{+0.12}_{-0.06}$ & $4.97^{+0.35}_{-0.62}$ \\
1718 & $3.27^{+0.09}_{-0.11}$ & $1.94^{+0.05}_{-0.05}$ & $16.66^{+0.02}_{-0.02}$ & $53.44^{+0.02}_{-0.02}$ & $-0.47^{+0.01}_{-0.01}$ & $48.69^{+0.02}_{-0.02}$ & $5.22^{+0.08}_{-0.07}$ \\
1858 & $2.85^{+0.10}_{-0.08}$ & $1.66^{+0.04}_{-0.04}$ & $16.69^{+0.02}_{-0.02}$ & $53.41^{+0.02}_{-0.02}$ & $-0.52^{+0.01}_{-0.02}$ & $48.67^{+0.02}_{-0.02}$ & $5.16^{+0.08}_{-0.08}$ \\
1970 & $1.97^{+0.13}_{-0.13}$ & $1.40^{+0.06}_{-0.06}$ & $16.81^{+0.04}_{-0.03}$ & $53.49^{+0.05}_{-0.05}$ & $-0.68^{+0.03}_{-0.03}$ & $48.74^{+0.05}_{-0.05}$ & $4.65^{+0.14}_{-0.15}$ \\
2075 & $2.34^{+0.15}_{-0.13}$ & $1.18^{+0.05}_{-0.05}$ & $16.70^{+0.03}_{-0.03}$ & $53.32^{+0.05}_{-0.05}$ & $-0.59^{+0.03}_{-0.03}$ & $48.58^{+0.05}_{-0.05}$ & $5.09^{+0.13}_{-0.14}$ \\
2163 & $<2.4$ & $>1.05$ & $>16.67$ & $>53.25$ & $<-0.58$ & $>48.50$ & $<5.23$ \\
2458 & $<2.3$ & $>0.86$ & $>16.65$ & $>53.16$ & $<-0.59$ & $>48.42$ & $<5.38$ \\
\enddata

\tablecomments{Quoted uncertainties are $1\sigma$ asymmetric errors.}
\end{deluxetable*}
\endgroup

\section{Comparison TDEs sample}

Table~\ref{tab:radio_tde_sample} summarizes the comparison radio TDEs selected from the sample of \citealt{Zhou2026}. We retain only measurements obtained within $4.8\leq\nu\leq5.2\,\mathrm{GHz}$.

\begingroup
\startlongtable
\newcommand{\centerrefs}[1]{%
  \begin{tabular}[t]{@{}c@{}}#1\end{tabular}%
}
\begin{deluxetable*}{cccc}
\tablecaption{Properties and radio data references for the 5~GHz radio TDE comparison sample}\label{tab:radio_tde_sample}
\tabletypesize{\small}
\tablewidth{0pt}
\renewcommand{\arraystretch}{0.95}
\tablehead{
\colhead{Source} &
\colhead{Redshift} &
\colhead{\shortstack{Discovery time}} &
\colhead{\shortstack{Radio data references}}
}
\startdata
FIRST~J153350.8+272729 & 0.03243 & 1986 Nov 6--Dec 13 & \centerrefs{\citealt{Ravi2022}} \\
\tableline
Arp~299-B~AT1 & 0.010411 & 2005 Jan 30 & \centerrefs{\citealt{Mattila2018}} \\
\tableline
IGR~J12580+0134 & 0.00411 & 2011 Jan 6 & \centerrefs{\citealt{Irwin2015}\\\citealt{Yuan2016}\\\citealt{Perlman2017}\\\citealt{Perlman2022}} \\
\tableline
Swift~J1644+57 & 0.354 & 2011 Mar 25 & \centerrefs{\citealt{Berger2012}\\\citealt{Zauderer2013}\\\citealt{Eftekhari2018}\\\citealt{Cendes2021SwiftJ1644}} \\
\tableline
ASASSN-14ae & 0.0436 & 2014 Jan 25 & \centerrefs{\citealt{Cendes2024}} \\
\tableline
ASASSN-14li & 0.0206 & 2014 Nov 22 & \centerrefs{\citealt{Alexander2016}\\\citealt{Velzen2016}\\\citealt{Bright2018}\\\citealt{Anumarlapudi2024}} \\
\tableline
CNSS~J0019+00 & 0.018 & 2015 Mar 21 & \centerrefs{\citealt{Anderson2020}} \\
\tableline
ASASSN-15oi & 0.0484 & 2015 Aug 14 & \centerrefs{\citealt{Horesh2021ASASSN15oi}\\\citealt{Hajela2025}\\\citealt{Anumarlapudi2024}} \\
\tableline
PS16dtm & 0.0804 & 2016 Aug 12 & \centerrefs{\citealt{Cendes2024}} \\
\tableline
iPTF~16fnl & 0.016328 & 2016 Aug 26 & \centerrefs{\citealt{Horesh2021iPTF16fnl}\\\citealt{Cendes2024}} \\
\tableline
AT~2018zr & 0.071 & 2018 Mar 2 & \centerrefs{\citealt{Cendes2024}} \\
\tableline
AT~2018cqh & 0.048 & 2018 Jun 16 & \centerrefs{\citealt{Zhang2024AT2018cqh}\\\citealt{Yang2025}} \\
\tableline
AT~2018hco & 0.088 & 2018 Sep 18 & \centerrefs{\citealt{Horesh2018AT2018hco}\\\citealt{Cendes2024}} \\
\tableline
AT~2018hyz & 0.04573 & 2018 Oct 14 & \centerrefs{\citealt{Gomez2020}\\\citealt{Cendes2022}\\\citealt{Sfaradi2024}\\\citealt{Anumarlapudi2024}\\\citealt{Cendes2024}\\\citealt{Cendes2026AT2018hyz}} \\
\tableline
AT~2019ahk & 0.0262 & 2019 Jan 29 & \centerrefs{\citealt{Christy2024ASASSN19bt}\\\citealt{Anumarlapudi2024}} \\
\tableline
AT~2019dsg & 0.0512 & 2019 Apr 9 & \centerrefs{\citealt{Stein2021}\\\citealt{Cannizzaro2021}\\\citealt{Cendes2021a}\\\citealt{Mohan2022}\\\citealt{Cendes2024}} \\
\tableline
AT~2019ehz & 0.074 & 2019 Apr 29 & \centerrefs{\citealt{Cendes2024}} \\
\tableline
AT~2019eve & 0.081 & 2019 May 5 & \centerrefs{\citealt{Cendes2024}} \\
\tableline
AT~2019qiz & 0.01513 & 2019 Sep 18 & \centerrefs{\citealt{OBrien2019a}\\\citealt{OBrien2019b}\\\citealt{Anumarlapudi2024}\\\citealt{Alexander2026}} \\
\tableline
AT~2019teq & 0.0878 & 2019 Oct 20 & \centerrefs{\citealt{Cendes2024}\\\citealt{Zhou2026}} \\
\tableline
AT~2020mot & 0.070 & 2020 Jun 14 & \centerrefs{\citealt{Liodakis2023}\\\citealt{Cendes2024}} \\
\tableline
AT~2020opy & 0.159 & 2020 Jul 8 & \centerrefs{\citealt{Goodwin2023a}} \\
\tableline
eRASSt~J045650.3$-$203750 & 0.077 & 2020 Sep 8 & \centerrefs{\citealt{Liu2024}} \\
\tableline
AT~2020vdq & 0.045 & 2020 Oct 4 & \centerrefs{\citealt{Somalwar2025}\\\citealt{Zhou2026}} \\
\tableline
AT~2020vwl & 0.0325 & 2020 Oct 8 & \centerrefs{\citealt{Goodwin2023b}\\\citealt{Goodwin2025AT2020vwl}} \\
\tableline
eRASSt~J210858$-$562832 & 0.043 & 2020 Oct 26 & \centerrefs{\citealt{Goodwin2025}} \\
\tableline
AT~2020wjw & 0.1 & 2020 Nov 28 & \centerrefs{\citealt{Goodwin2024AT2020wjw}} \\
\tableline
eRASSt~J011431$-$593654 & 0.16 & 2020 Nov 24 & \centerrefs{\citealt{Goodwin2025}} \\
\tableline
AT~2022cmc & 1.19325 & 2022 Feb 11 & \centerrefs{\citealt{Andreoni2022}\\\citealt{Pasham2023}\\\citealt{Rhodes2023}} \\
\tableline
AT~2020afhd & 0.027 & 2024 Jan 1 & \centerrefs{\citealt{Wang2025}\\\citealt{Christy2024AT2020afhd}} \\
\tableline
AT~2024tvd & 0.04494 & 2024 Aug 25 & \centerrefs{\citealt{Sfaradi2025}} \\
\tableline
EP250702a & 1.036 & 2025 Jul 2 & \centerrefs{\citealt{Li2026EP250702a}\\\citealt{Levan2025EP250702a}} \\
\enddata
\end{deluxetable*}
\endgroup

\section{Off-axis jet model}

Table~\ref{tab:vegas_priors} lists the prior ranges and posterior constraints for the off-axis jet fits described in Sect.~\ref{subsec:jet}. 

\begingroup
\begin{deluxetable*}{cccc}
\tablecaption{Prior and posterior constraints for the off-axis jet model fits}\label{tab:vegas_priors}
\tabletypesize{\small}
\tablewidth{0pt}
\tablehead{
\colhead{Parameter} &
\colhead{Prior range} &
\multicolumn{2}{c}{Posterior} 
}
\startdata
$t_\mathrm{0}$ & Fixed & $58803.5$ & $59335.0$ \\
$\log_{10}E_{\rm iso}$ & $\left[50,\,60\right]$ & $56.10^{+0.53}_{-0.55}$ & $54.40^{+0.85}_{-0.96}$ \\
$\log_{10}\Gamma_0$ & $[1,\,4]$ & $3.24^{+0.46}_{-0.58}$ & $2.71^{+0.75}_{-0.83}$ \\
$\theta_{\rm c}$ & $[0.001,\,0.5]$ & $0.01\pm0.01$ & $0.05^{+0.04}_{-0.01}$ \\
$\theta_{\rm v}$ & $[0,\,\pi/2]$ & $0.39^{+0.05}_{-0.04}$ & $0.35^{+0.50}_{-0.08}$ \\
$\log_{10}A_\star$ & $[-2,\,2]$ & $-1.38^{+0.52}_{-0.43}$ & $0.31^{+1.31}_{-1.11}$ \\
$k_{\rm m}$ & $[0.5,\,4.0]$ & $0.51\pm0.01$ & $1.47^{+0.42}_{-0.22}$ \\
$p$ & $[2.01,\,4.0]$ & $3.25\pm0.07$ & $3.26^{+0.12}_{-0.15}$ \\
$\log_{10}\epsilon_{\rm e}$ & $[-3,\,-0.3]$ & $-1.99^{+0.83}_{-0.66}$ & $-1.75^{+0.52}_{-0.76}$ \\
$\log_{10}\epsilon_{\rm B}$ & $[-6,\,-0.3]$ & $-4.85^{+0.98}_{-0.77}$ & $-3.63^{+2.36}_{-1.43}$ \\
$\log_{10}\xi_{\rm e}$ & $[-3,\,0]$ & $-1.66^{+0.63}_{-0.56}$ & $-1.84^{+0.52}_{-0.62}$ \\
\enddata
\end{deluxetable*}
\endgroup

\clearpage
\twocolumngrid
\section{Numerical Simulations}\label{sec:simu}
Through simulations, the resulting distributions of cosmic-ray electron energy density and magnetic field enable modeling of the associated synchrotron emission. 
Prior to the radiative transfer calculations, we transform the data from the two-dimensional spherical coordinates used in the simulation to a cylindrical coordinate system. This transformation facilitates radiative transfer calculations along the polar direction and allows us to derive the radio spectra.

\begin{figure}
\centering
\includegraphics[width=0.3\columnwidth]{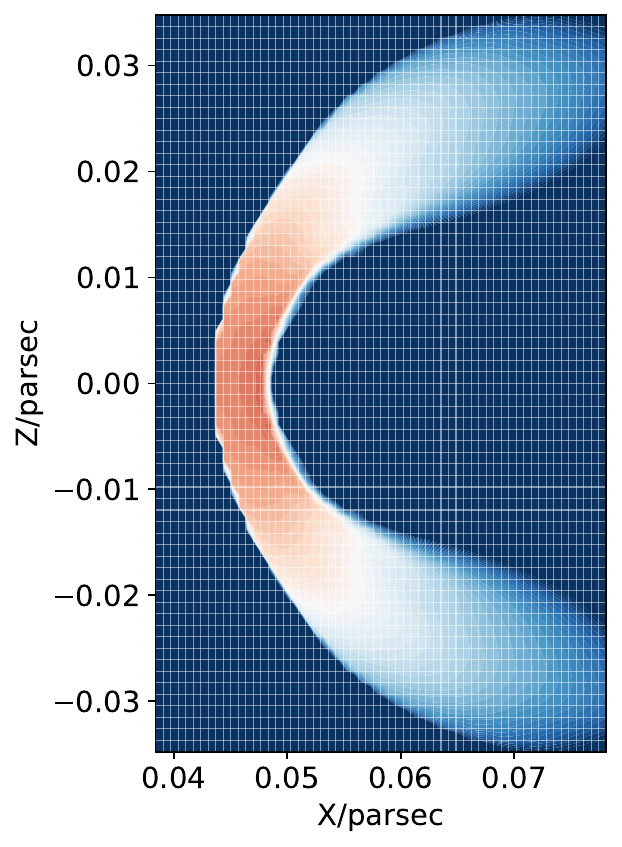}
\caption{Distribution of $e_2$ in the cylindrical coordinate system after coordinate and grid transformation. }
\label{figA1}
\end{figure}

To improve computational efficiency, we apply this coordinate and grid transformation only to the bow-shock region (i.e., the radiative zone enclosed by the bow shock) using an interpolation method, as illustrated in Fig.~2 in \cite{Mou_2026}. 
The cylindrical grid consists of 64 zones along the vertical ($Z$) axis and 60 zones in the radial direction. 
Fig.~\ref{figA1} shows the transformed data for 2023-Oct-24 ($\Delta t=1718$ d). 

Under the assumption that physical quantities remain constant within an individual grid cell, the variation of the specific intensity $I_\nu$ along a ray segment of length $\Delta z$ is governed by: $\Delta I_{\nu}/\Delta z=-\alpha_{\nu} I_{\nu}+j_{\nu}$, where $j_{\nu}$ represents the synchrotron volume emissivity and $\alpha_{\nu}$ is the corresponding absorption coefficient. 

The formal solution of the radiative transfer equation across a single, homogeneous grid cell yields the specific intensity at the downstream boundary. Denoting this intensity as $I_{\nu}(i+1)$, it is related to the upstream intensity $I_{\nu}(i)$ by:
\begin{equation}
I_{\nu}(i+1)=I_{\nu}(i) e^{-\Delta \tau_{\nu}(i)} + S_{\nu}(i)\left[1-e^{-\Delta \tau_{\nu}(i)}\right].
\end{equation}
Here, $S_{\nu}(i) = j_{\nu}(i)/\alpha_{\nu}(i)$ is the source function in the $i$-th cell, and $\Delta \tau_{\nu}(i)=\alpha_{\nu}(i) \Delta x_i $ is the corresponding optical depth. Starting from an initial background intensity at the first cell, this solution is applied iteratively along the entire line of sight to obtain the emergent intensity $I_{\nu, N}$ from the final ($N$-th) cell at the far boundary of the emitting region.

The total spectral flux density \( F_{\nu} \) is then obtained by integrating the emergent intensity $I_{\nu, N}$ over the projected surface of the radiative zone facing the observer:
$F_{\nu} = \int_{\text{surface}} I_{\nu, N} \, \cos \theta \, d\Omega, $
where $\theta$ is the angle between the local surface normal and the observer's line of sight, and $d\Omega$ is the solid angle subtended by a surface element at the source distance.

For the polar direction, the flux is calculated by summing over cylindrical annuli:
\begin{equation}
F_{\nu} = \frac{1}{d^2_L} \sum_{R} I_{\nu, N}(R) \cdot 2\pi R \Delta R ,
\end{equation}
where $d_L$ represents the luminosity distance of AT2019avd. 

After exploring a range of parameter sets, we selected a representative case that reproduces the observed spectra for presentation in the main text. Fig.~\ref{figA3} compares the synthetic spectra produced by three distinct models. The spectral analysis indicates that reducing the cloud size leads to a more compact emitting region, which shifts the synchrotron self-absorption peak to higher frequencies. As the cloud size increases, however, the peak frequency becomes relatively insensitive to further changes in cloud size. A larger $R_c$ requires a lower mass outflow rate to recover a similar radio flux. In this case, the increase in the sight-line length of the radiative zone is offset by the simultaneous decline in relativistic electron density and magnetic field strength, resulting in only a marginal variation in the synchrotron optical depth.  

\begin{figure*}
\centering
\includegraphics[width=0.98\textwidth]{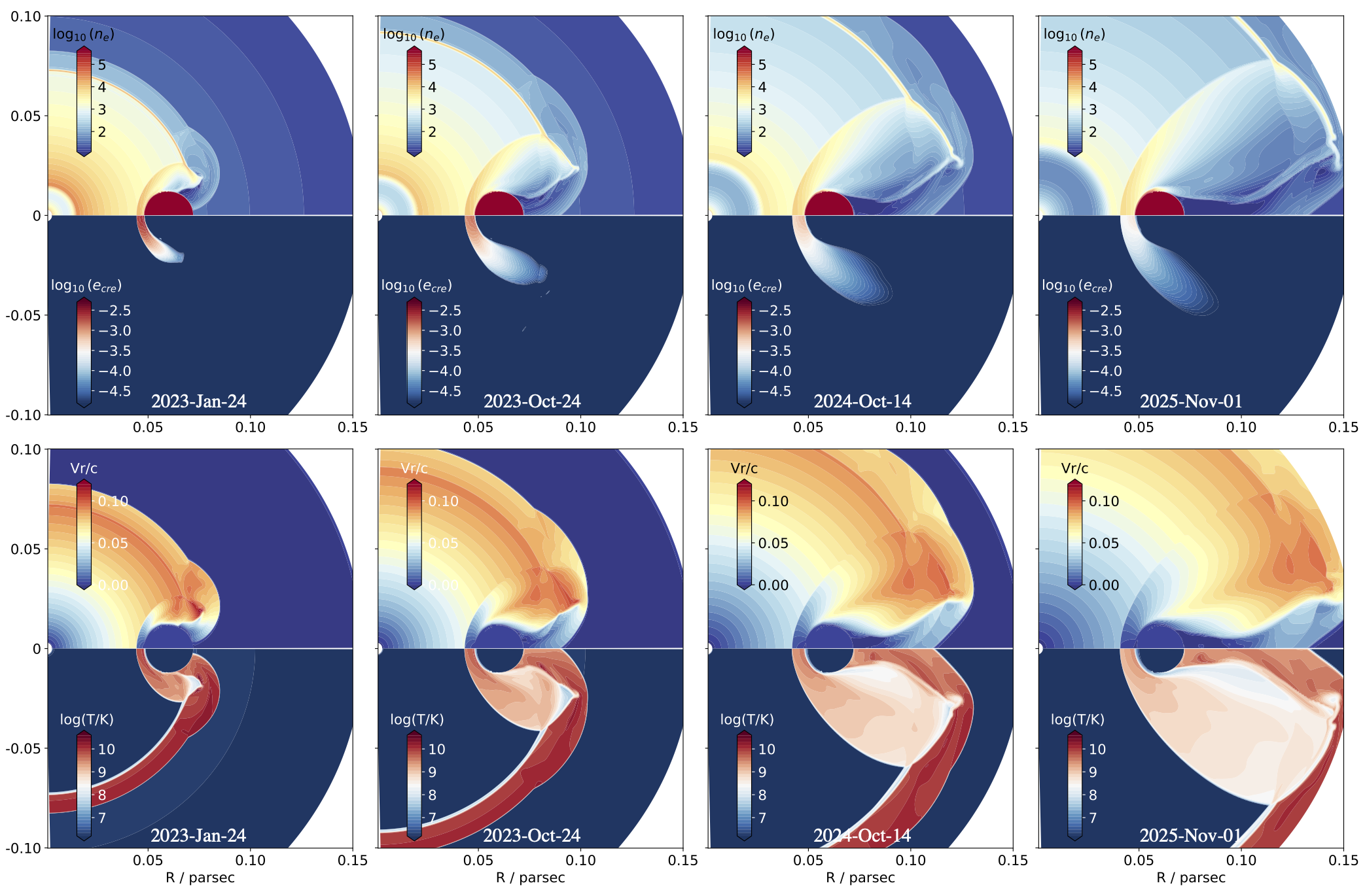}
\caption{Temporal evolution of outflow-cloud interactions. The upper panels show the density distribution and the energy density of cosmic-ray electrons. The lower panels show the radial velocity (in unit of light speed) and temperature. }
\label{figA2}
\end{figure*}

\begin{figure*}
\centering
\includegraphics[width=0.31\textwidth]{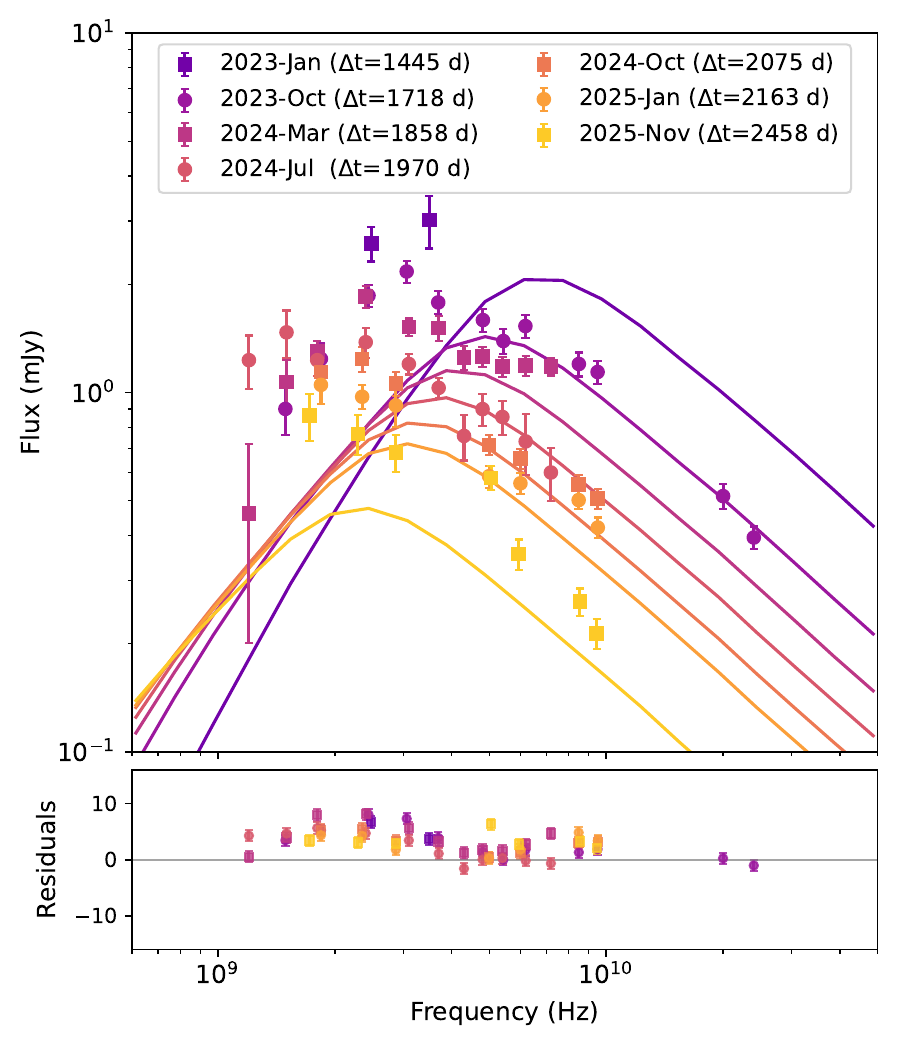}
\includegraphics[width=0.31\textwidth]{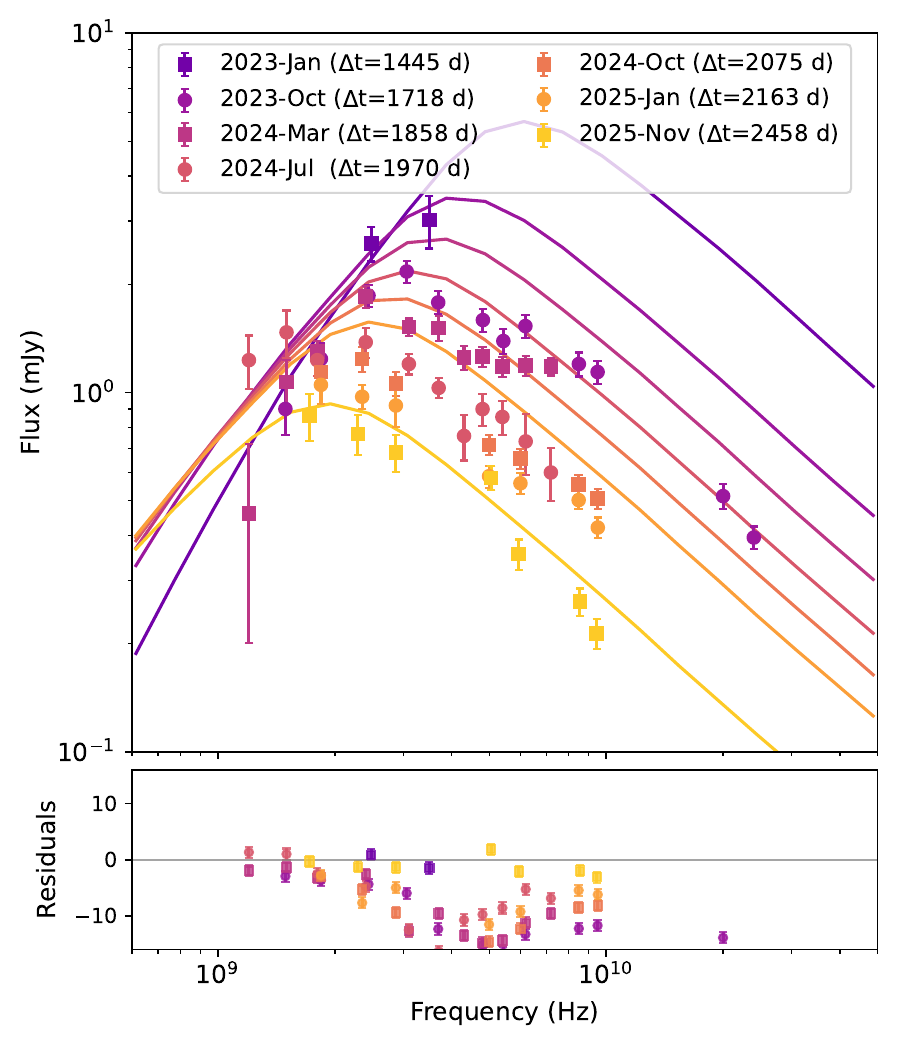}
\includegraphics[width=0.31\textwidth]{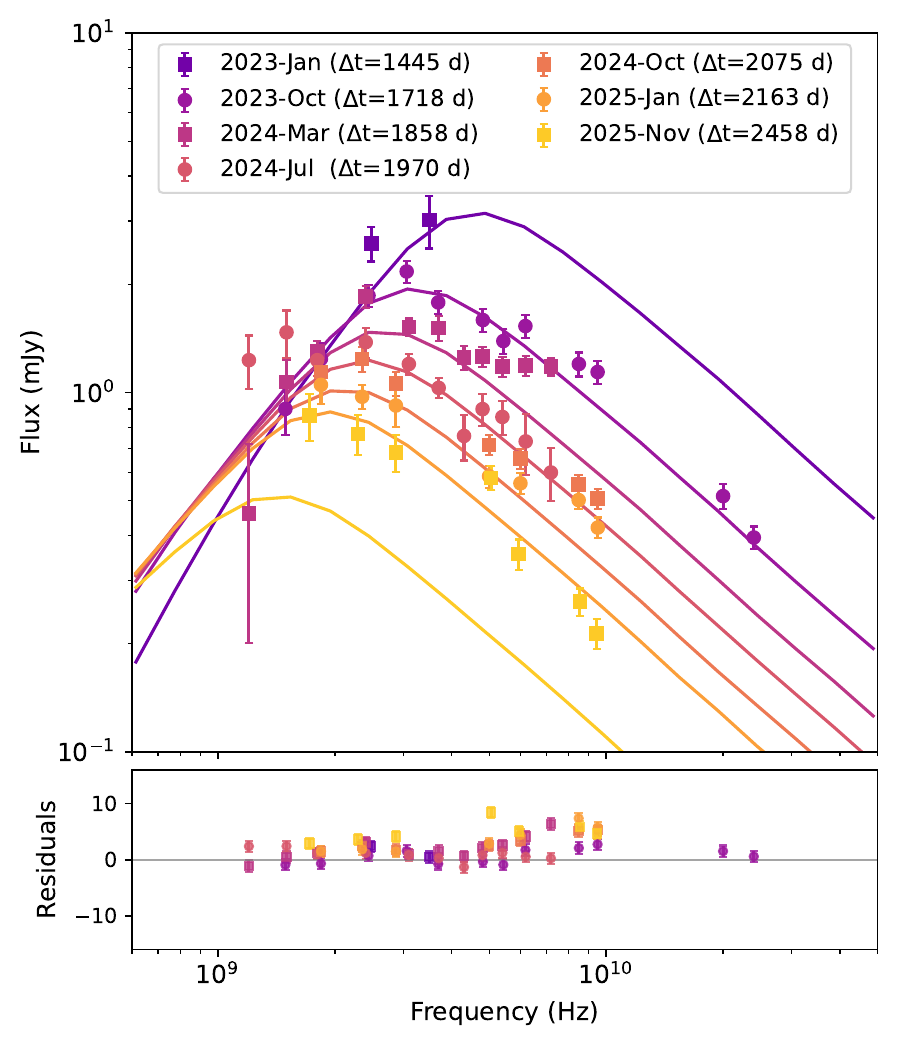}
\caption{Synthetic radio spectra for models with different cloud distances ($d_\mathrm{c}$) and sizes ($R_\mathrm{c}$). The left panel displays a model in which $d_\mathrm{c}$ and outflow velocity are reduced simultaneously ($d_\mathrm{c} = 0.036$ pc; velocity decreasing linearly from $0.05c$ to $0.0125c$), while all other parameters remain unchanged ($R_\mathrm{c} = 0.012$ pc and $\dot{M}_{\rm out} = 0.17 ~\msunyr$). The middle panel illustrates the effect of increasing only the cloud size to $R_\mathrm{c} = 0.02$ pc (relative to the model presented in the main text with $R_\mathrm{c}=0.012$ pc). The right panel presents a model with $R_\mathrm{c} = 0.02$ pc that reproduces the observations by scaling the mass outflow rate to $\dot{M}_{\rm out} = 0.11 ~\msunyr$ through adjusting the outflow density alone.  }
\label{figA3}
\end{figure*}

\end{document}